\documentclass{article}

\usepackage{arxiv}

\usepackage[utf8]{inputenc} 
\usepackage[T1]{fontenc}    
\usepackage{hyperref}       
\usepackage{url}            
\usepackage{booktabs}       
\usepackage{amsthm,amsmath,amsfonts,amssymb} 
\usepackage{nicefrac}       
\usepackage{microtype}      
\usepackage{cleveref}       
\usepackage{lipsum}         
\usepackage{graphicx}
\usepackage{natbib}
\usepackage{doi}
\usepackage{multirow}

\title{Understanding and Using the Relative Importance Measures Based on Orthogonalization and Reallocation}

\newif\ifuniqueAffiliation
\uniqueAffiliationtrue

\ifuniqueAffiliation 
\author{ \href{https://orcid.org/0009-0002-4187-513X}{\includegraphics[scale=0.06]{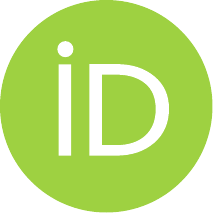}\hspace{1mm}Tien-En Chang} \\
	Industrial Engineering\\
	National Taiwan University\\
	Taipei, Taiwan \\
	\texttt{f09622009@ntu.edu.tw} \\
	\And
	\href{https://orcid.org/0000-0002-7951-9950}{\includegraphics[scale=0.06]{orcid.pdf}\hspace{1mm}Argon Chen}\thanks{Corresponding author.} \\
	Industrial Engineering\\
	National Taiwan University\\
	Taipei, Taiwan \\
	\texttt{achen@ntu.edu.tw} \\
}

\date{}

\renewcommand{\shorttitle}{ORM Relative Importance}

\begin{document}
\maketitle

\begin{abstract}
A class of relative importance measures based on orthogonalization and reallocation, ORMs, has been found to effectively approximate the General Dominance index (GD). In particular, Johnson's Relative Weight (RW) has been deemed the most successful ORM in the literature. Nevertheless, the theoretical foundation of the ORMs remains unclear. To further understand the ORMs, we provide a generalized framework that breaks down the ORM into two functional steps: orthogonalization and reallocation. To assess the impact of each step on the performance of ORMs, we conduct extensive Monte Carlo simulations under various predictors' correlation structures and response variable distributions. Our findings reveal that Johnson's minimal transformation consistently outperforms other common orthogonalization methods. We also summarize the performance of reallocation methods under four scenarios of predictors’ correlation structures in terms of the first principal component and the variance inflation factor (VIF). This analysis provides guidelines for selecting appropriate reallocation methods in different scenarios, illustrated with real-world dataset examples. Our research offers a deeper understanding of ORMs and provides valuable insights for practitioners seeking to accurately measure variable importance in various modeling contexts.
\end{abstract}

\keywords{Relative importance analysis \and Johnson’s relative weight \and General dominance \and Shapley value \and Monte Carlo Simulation}
\section{Introduction}
Regression analysis is widely used to build predictive models. Beyond prediction, researchers and practitioners are often interested in the importance of the predictors within the model. Therefore, finding a proper measure of the relative importance of predictors in linear models has been of interest in the literature. In this context, one of the most well-known definitions of the relative importance in a linear model is the proportionate contribution each predictor makes to explaining the variance of the response variable by considering both its direct effect and its effect when combined with the other predictors~\citep{johnson2004history}.

Traditional measures such as simple correlations and regression coefficients were initially considered but soon found to have serious disadvantages when predictors are correlated. General Dominance index
(GD) addresses these issues by providing a comprehensive and theoretically meaningful measure of relative importance~\citep{budescu1993dominance}. Therefore, GD is generally regarded as the most plausible measure of
relative importance. For a detailed review of various relative importance measures, see~\cite{gromping2015variable}.

However, GD is computationally intractable by its definition. \emph{The Orthogonalization-Reallocation Measures (ORMs)}, which we will introduce in the following section, appear to be a viable and fast alternative. Despite their different mathematical formulations, empirical results show convergence between the ORMs and GD~\citep{chao2008quantifying, johnson2000heuristic, lebreton2004monte}. Among these, Relative Weight (RW) proposed by~\cite{johnson2000heuristic} is the most commonly used ORM due to its practical success. As a result, RW has been extended to various applications, such as logistic regression~\citep{tonidandel2010determining} and multivariate regression~\citep{lebreton2008multivariate, hong2012dominance}. Research
on determining the statistical significance of RW is detailed in~\cite{tonidandel2009determining} and its application to higher-order regression models is discussed by~\cite{lebreton2013residualized}. In addition, some research extends RW to the high-dimensional setting, where the number of predictors exceeds the sample size, and finds that RW is competitive with methods like lasso~\citep{tibshirani1996regression} and elastic net~\citep{zou2005regularization} in variable selection tasks~\citep{shen2020comprehensive, shen2021many}.

While RW has demonstrated empirical success, its theoretical underpinnings remain limited. Some researchers \citep{thomas2014johnson} have argued that the theoretical formulation of RW is flawed, suggesting
that RW should no longer be used despite its empirical success. This theoretical controversy surrounding RW highlights the need for a clearer framework for understanding ORMs and for a thorough performance analysis. Our research aims to address this gap by providing a framework and conducting a comprehensive evaluation of ORMs.

In this article, we first revisit GD and the existing ORMs. Next, we provide a general framework for the ORMs, which consists of two critical steps: orthogonalization and reallocation. To understand the
performance of ORMs, we conduct comprehensive Monte Carlo simulations and summarize the key factors. Finally, we offer guidelines on how to use them effectively. We also demonstrate our findings with real-world dataset examples.
\section{Two Types of the Relative Importance Measures}
In this section, we briefly review general dominance index and the measures based on orthogonalization of the predictors.
\subsection{Notation and Setup}
Given that $X=[x_1,\ldots,x_p ]$ is an $n\times p$ matrix that contains $p$ predictors, $\beta$ is a $p\times 1$ vector of regression coefficients, $y$ is an $n\times 1$ vector of response variable and $\epsilon$ is an $n\times 1$ noise vector that is orthogonal to the columns of $X$. Assume that $X$ has full column rank and all $x_i$ and $y$ are standardized to zero mean and unit norm (length). A linear regression model is defined as follows:
\begin{equation}
    y = X\beta + \epsilon.
    \label{eq: linear regression}
\end{equation}
In addition, let the singular value decomposition (SVD) of $X$ be
\begin{equation}
    X=USV^\top,
    \label{eq: svd}
\end{equation}
where $U$ is an $n\times p$ left singular matrix, $S$ is a $p\times p$ diagonal matrix with positive singular values, and $V$ is a $p\times p$ right singular matrix. When the transformed predictors are involved in the equation, we use indices $i$ to indicate the original predictors, and $j$ to indicate the transformed predictors.
\subsection{General Dominance Index (GD) / Shapley Value}
Among various measures of relative importance, General Dominance index (GD)~\citep{budescu1993dominance} stands out as one of the most plausible methods. GD evaluates a predictor’s importance by averaging its contribution in predicting $y$ across all possible combinatorial sub-models, providing a comprehensive perspective. Specifically, the squared multiple correlation $R_{y\cdot X}^2$ is used to assess the contribution when all predictors $X$ are used in predicting $y$. Therefore, GD of $x_i$ (with respect to $y$) can be defined as the weighted average of all possible increments:
\begin{equation}
    \mathrm{GD}(x_i)=\frac{1}{p}\sum_{S\subseteq P\setminus\{i\}}\frac{1}{\binom{p-1}{|S|}}\left[R_{y\cdot X_{S\cup \{i\}}}^2-R_{y\cdot X_S}^2\right]
    \label{eq: general dominance}
\end{equation}
where $P\in[1,\ldots,p]$ is a predictor index set, $S$ is a subset of predictor indices, $|\cdot|$ denotes the cardinality of a set and $S\subseteq P\setminus\{i\}$ denotes all possible subsets excluding the index of predictor $i$ and $X_S$ denotes a subset of predictors with indices $S$. For example, when $p=3$, the general dominance of $x_1$ is
\begin{align}
    \mathrm{GD}(x_1) &= \frac{1}{3}\Biggl( R_{y\cdot x_1}^2 + \frac{1}{2}\Bigl[ \bigl(R_{y\cdot x_1x_2}^2 - R_{y\cdot x_2}^2\bigr) +  \bigl(R_{y\cdot x_1x_3}^2 - R_{y\cdot x_3}^2\bigr) \Bigr] + R_{y\cdot x_1x_2x_3}^2 - R_{y\cdot x_2x_3}^2 \Biggr).
    \label{eq: GD p=3}
\end{align}
It is worth noting that the concept of GD coincides with the Shapley value~\citep{shapley1953game}, which originates from game theory and is recently used in the context of explainable machine learning. However, GD becomes computationally intractable as the number of predictors increases because it requires calculations for $2^p-1$ sub-models.
\subsection{The Orthogonalization-Reallocation Measures (ORMs)}
The three measures introduced in this subsection share a common idea: estimating variable importance through the orthogonalization transformation of the original predictors. For generalization purposes, we refer to these measures as the Orthogonalization-Reallocation Measures (ORMs). The emphasis on orthogonalization arises from the requirement that the orthonormal vectors $\tilde{z}_1,\ldots,\tilde{z}_p$, must lie within the subspace spanned by the original predictors. Thus, we must always have a relation $\tilde{Z}=XT$ where $T\in\mathbb{R}^{p\times p}$ is a linear transformation such that $\tilde{Z}^\top\tilde{Z}=I_p$.

Although there are infinitely many $\tilde{Z}$ that satisfy such criterion, the most popular choice is the minimal transformation proposed by~\cite{johnson1966minimal}, which minimizes the sum of squared differences between $Z$ and $X$. This can be formulated by minimizing the objective function:
\begin{equation}
    O_J(\tilde{Z})=\mathrm{Tr}\left((\tilde{Z}-X)^\top(\tilde{Z}-X)\right),
    \label{eq: obj minimal trans.}
\end{equation}
where $\mathrm{Tr}(\cdot)$ denotes the trace of a matrix. The minimizer of this problem is given by
\begin{equation}
    Z=X(X^\top X)^{-1/2}=UV^\top,
    \label{eq: minimal trans.}
\end{equation}
which we refer to as Johnson’s $Z$ in this paper.
\subsubsection{The R.M. Johnson's (1966) Method}
Based on his proposed orthonormality transformation, R.M. Johnson suggested that the importance of $x_i$ can be simply obtained from the importance of its corresponding orthogonal transformation $z_i$:
\begin{equation}
    w_i^2=(z_i^\top y)^2=\rho_{z_iy}^2,
    \label{eq: importance weight}
\end{equation}
the squared correlation between $y$ and $z_i$, because $Z$ is closest to $X$ in the least squares sense, and the proportion of explained variance can be decomposed by orthogonal predictors $Z$, i.e., $\sum_{i=1}^p w_i^2=R^2_{y\cdot X}$. However, when the original predictors tend to be highly correlated, the angles between $z_i$ and $x_i$ become large. Therefore, the importance of orthogonal predictors $w_i$ may not be able to accurately represent that of the original predictors (See Figure 1 in~\cite{green1978new, thomas2014johnson}). (The CAR score~\citep{zuber2010high} coincides with $w_i$.)
\subsubsection{The Green, Carroll and DeSarbo (GCD) Method}
To address the limitation in Johnson (1966),~\cite{green1978new} proposed relating the importance of $z_1,\ldots,z_p$ back to $x_1,\ldots,x_p$ using the normalized squared regression coefficients of $Z$ on $X$. Specifically, the regression coefficients are given by
\begin{equation}
    \Gamma_Z=(X^\top X)^{-1}X^\top Z=(X^\top X)^{-1/2}=VS^{-1}V^\top,
    \label{eq: regress z on x}
\end{equation}
where $\gamma_{Z,ij}$ denotes the element of $\Gamma_Z$. The normalized squared regression coefficient of $z_j$ on $x_i$ is defined as
\begin{equation}
    \gamma_{ij}^{*2}=\frac{\gamma_{Z,ij}^2}{\sum_{i=1}^p \gamma_{Z,ij}^2}.
    \label{eq: RegPA with minimal trans}
\end{equation}
As a result, the importance of each $x_i$ is obtained by
\begin{equation}
    \delta_i^2=\sum_{j=1}^p \gamma_{ij}^{*2}\rho_{z_jy}^2.
    \label{eq: GCD}
\end{equation}
Although GCD provides a method to connect the importance of orthogonal predictors to the original predictors, the use of normalized regression coefficients has been criticized due to the inter-correlation between $x_i$~\citep{johnson2000heuristic}. 
\subsubsection{The J.W. Johnson's (2000) Method / Relative Weight (RW)}
\cite{johnson2000heuristic} proposed an alternative method that relates the importance back by regressing $X$ on $Z$:
\begin{equation}
    L_Z=(Z^\top Z)^{-1}Z^\top X=Z^\top X=VSV^\top.
    \label{eq: correlation between minimal trans and x}
\end{equation}
The element $\ell_{Z,ij}$ of $L_Z$ represents the simple correlation between $z_j$ and $x_i$. J.W. Johnson claimed that using $\ell_{Z,ij}^2$ in place of $\gamma_{ij}^{*2}$ avoided the problem of correlated variables introduced by $\delta^2$. By replacing $\gamma_{ij}^{*2}$ in Equation~\eqref{eq: GCD}, we obtain the importance of $x_i$:
\begin{equation}
    \varepsilon_i = \sum_{j=1}^p \ell_{Z,ij}^2\rho_{z_jy}^2,
    \label{eq: RW}
\end{equation}
which is referred to as Johnson's relative weight (RW). To demonstrate the performance of RW, J.W. Johnson considered GD as the benchmark and showed the mean absolute deviations between RW and GD are smaller than those between GCD and GD in 28 of the 31 real and simulated datasets. In addition, due to its computational efficiency, RW can be seen as an alternative to GD when the number of predictors is large. Subsequent empirical studies also support this observation~\citep{chao2008quantifying, lebreton2004monte}, making RW a favorable method for assessing variable relative importance. (RW coincides with the method proposed by~\cite{genizi1993decomposition}.)

In the next section, we begin by introducing a general framework of the ORMs and provide insights based on this framework. We then apply these insights in comprehensive Monte Carlo simulations to enhance our understanding of the ORMs.
\section{How do the ORMs work?}
\begin{figure}[htb]
\includegraphics[width=\textwidth]{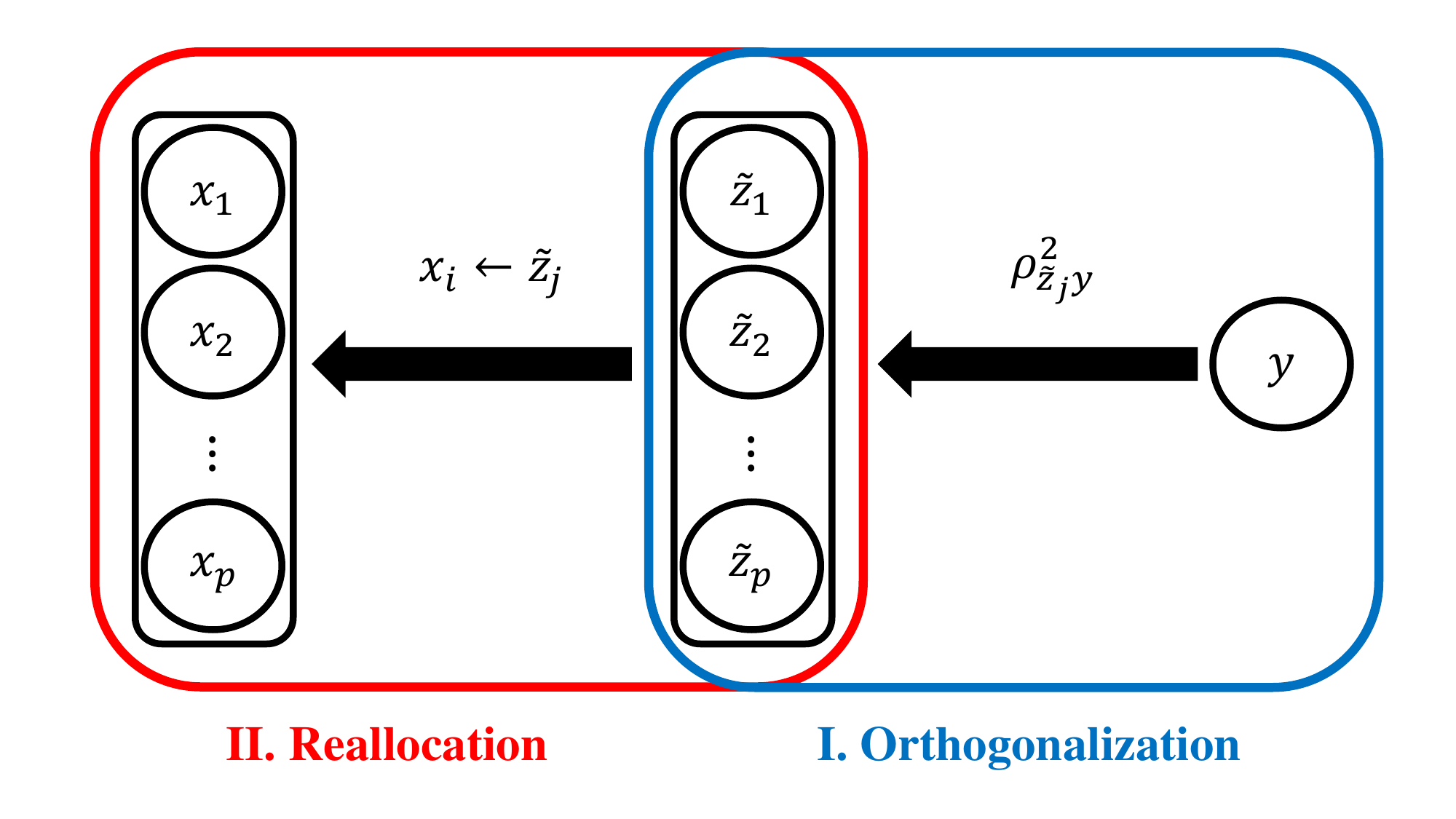}
\caption{The general framework of the ORMs.}
\label{fig: framework of ORM}
\end{figure}

In this section, we generalize the framework first mentioned in~\cite{wallard2019grouping}, which highlights the two key components of the ORMs: Orthogonalization and Reallocation (see Figure~\ref{fig: framework of ORM}).
First, ORMs perform orthogonalization on the original predictors $X$ and calculate the contribution of the orthogonal predictors $\tilde{Z}$ toward explaining the variance in the response variable $y$. Since $\tilde{z}_1,\ldots,\tilde{z}_p$ are orthogonal to each other, their contributions are represented by the squared simple correlations with the response variable $\rho_{\tilde{z}_jy}^2$. The importance of $\tilde{z}_1,\ldots,\tilde{z}_p$ is then based on their contributions. 

Next, the importance of $\tilde{z}_1,\ldots,\tilde{z}_p$ is reallocated back to the original predictors X to estimate the relative importance of each $x_i$. The reallocation can be expressed as a matrix $A^{\tilde{Z}}$ with the element $a_{ij}^{\tilde{Z}}$ representing the proportion of $\tilde{z}_j$ importance allocated to $x_i$. Consequently, given the reallocation matrix $A^{\tilde{Z}}$ and the orthogonal predictors $\tilde{Z}$, the relative importance of $x_i$ estimated by the ORM is denoted by $D_{A,i}^{\tilde{Z}}$ and can be formulated as:
\begin{equation}
    D_{A,i}^{\tilde{Z}}=\sum_{j=1}^p a_{ij}^{\tilde{Z}}\rho_{\tilde{z}_jy}^2.
    \label{eq: ORM}
\end{equation}
\subsection{The Orthogonalization}
All the ORMs introduced in the previous section use Johnson's minimal transformation for orthogonalization. However, as noted by~\cite{johnson2000heuristic}, there are an infinite number of orthogonalization methods, each resulting in different predictor importance. To further understand how orthogonalization affects the ORMs, we first generalize the use of orthogonalization and then compare them through comprehensive Monte Carlo simulations. Besides Johnson’s transformation, below are some common orthogonalization methods:
\begin{itemize}
    \item The Gram-Schmidt Orthogonalization: Without loss of generality, let us define the corresponding orthogonal predictors $Z_{\mathrm{GS}}$ as those obtained by applying the Gram-Schmidt process to the original predictors $X$ in ascending order.
    \item The Standardized Principal Components: The standardized principal components can be simply written as $Z_{\mathrm{PC}}=U$.
    \item The Varimax Method: Denote the loading matrix of X on an arbitrary orthogonal counterpart $\tilde{Z}$ as $L_{\tilde{Z}}=X^\top\tilde{Z}$. The Varimax predictors $Z_{\mathrm{VM}}$ can be obtained by maximizing the objective function proposed by~\cite{kaiser1958varimax}:
\end{itemize}
\begin{equation}
    O_{\mathrm{VM}}(\tilde{Z})=\sum_{j=1}^p\left(\frac{1}{p}\sum_{i=1}^p(\ell_{\tilde{Z},ij})^2-\frac{1}{p^2}\left(\sum_{i=1}^p \ell_{\tilde{Z},ij}\right)^2\right).
    \label{eq: varimax}
\end{equation}
\subsection{The Reallocation}
Over the past 50 years, researchers have focused on improving the reallocation while consistently using Johnson's $Z$. Many have proposed new reallocation methods to address the shortcomings of previous ones. Based on the framework in Equation~\eqref{eq: ORM}, the reallocations of the three ORMs introduced in the previous section, given an arbitrary $\tilde{Z}$, can be generalized as follows:
\begin{itemize}
    \item Reallocation of the R.M. Johnson’s (1966) method with arbitrary $\tilde{Z}$ is then:
    \begin{align}
        a_{ij}^{\tilde{Z}}=\mathrm{IdA}_{ij}^{\tilde{Z}}=\begin{cases}
            1, &\text{if }i=j, \\
            0, &\text{otherwise.}
        \end{cases}
        \label{eq: IdA}
    \end{align}
    In the rest of this paper, we refer to it as \emph{the identity reallocation (IdA)}. Thus, we write the ORM that uses this reallocation as $D_{\mathrm{IdA},i}^{\tilde{Z}}$.

    \item Reallocation of the Green, Carroll and DeSarbo (GCD) Method with arbitrary $\tilde{Z}$ is:
    \begin{align}
        a_{ij}^{\tilde{Z}}=\mathrm{RegPA}_{ij}^{\tilde{Z}}=\frac{\gamma^2_{\tilde{Z},ij}}{\sum_{i=1}^p \gamma^2_{\tilde{Z},ij}}.
        \label{eq: RegPA}
    \end{align}
    We name this reallocation \emph{the regression-based proportional reallocation (RegPA)} as it uses the proportion of squared regression coefficients. Thus, we write the ORM that uses this reallocation as $D_{\mathrm{RegPA},i}^{\tilde{Z}}$.

    \item Reallocation in the J. W. Johnson’s (2000) method with arbitrary $\tilde{Z}$ can be formulated as:
    \begin{align}
        a_{ij}^{\tilde{Z}}=\mathrm{CorPA}_{ij}^{\tilde{Z}}=\frac{\ell^2_{\tilde{Z},ij}}{\sum_{i=1}^p \ell^2_{\tilde{Z},ij}}.
        \label{eq: CorPA}
    \end{align}
    Since this reallocation uses the proportion of squared simple correlation coefficients, we name it \emph{the correlation-based proportional reallocation (CorPA)}. Thus, we write the ORM that uses this reallocation as $D_{\mathrm{CorPA},i}^{\tilde{Z}}$. It should be noted that when $\tilde{Z}$ is chosen to be Johnson’s $Z$, we have $\sum_{j=1}^p\ell_{Z,ij}^2=\sum_{i=1}^p\ell_{Z,ij}^2=1$ due to the symmetric $L_Z$ resulting from Johnson’s $Z$, i.e. $Z^\top X=X^\top Z$. Consequently, Equation~\eqref{eq: CorPA} can be simplified to $a_{ij}^{Z}=\ell_{Z,ij}^2$ for J. W. Johnson’s RW in Equation~\eqref{eq: RW}.
\end{itemize}
We find these reallocation methods share the following properties: (1) Non-negativity: $a_{ij}^{\tilde{Z}}\geq 0, \forall i,j=1,\ldots,p$. (2) Column-sum to one: $\sum_{i=1}^p a_{ij}^{\tilde{Z}}=1, \forall j=1,\ldots,p$. The first property, non-negativity, is generally considered a basic constraint since negative reallocation of importance is not logically interpretable. The second property, column-sum to one, ensures that $\sum_{i=1}^p(\sum_{j=1}^p a_{ij}^{\tilde{Z}}\rho^2_{\tilde{z}_jy})=\sum_{j=1}^p\rho^2_{\tilde{z}_jy}=R^2$. Different reallocation methods represent different perspectives on how the original predictors contribute to the importance of orthogonalized predictors. 

The critique by~\cite{thomas2014johnson} is based on interpretation of regressing $X$ on $Z$ in calculation of RW. However, we argue here that RW is best understood within the ORM framework. From the perspective of the framework, RW is simply employing Johnson's $Z$ for orthogonalization and CorPA as its reallocation method. RW, as well as GCD, can be seen as one of many valid ORM options for approximating GD through specific orthogonalization and reallocation approaches.

In general, the reallocation matrix $A^{\tilde{Z}}$ can be any matrix that satisfies the aforementioned properties. However, one may ask: “Does an optimal reallocation method exist?” Instead of finding an arbitrary reallocation matrix, we propose the most plausible yet computationally intractable reallocation solution is reallocation based on GD of $X$ in predicting $\tilde{Z}$:
\begin{equation}
    a_{ij}^{\tilde{Z}}=\mathrm{GD}_{\tilde{z}_j}(x_i)
    \label{eq: GDA}
\end{equation}
where $\mathrm{GD}_{\tilde{z}_j}(x_i)$ denotes general dominance index of $x_i$ in predicting $\tilde{z}_j$. This reallocation is the most plausible because not only it meets the properties mentioned above but also, in particular, best represents the contribution of $x_i$ in explaining $\tilde{z}_j$. We use this reallocation as a benchmark for other reallocation methods and name it \emph{the GD-based reallocation (GDA)}. It is important to note that GDA is as computationally intractable as GD, and is thus not a practical reallocation method when the number of predictors is large. However, it serves as a benchmark that allows us to gain new insights into ORMs. We denote the ORM using this reallocation as $D_{\mathrm{GDA},i}^{\tilde{Z}}$.
\subsection{The Need for a More Comprehensive Simulation}
When the number of predictors is two and Johnson’s $Z$ is used, it is easy to show that $\mathrm{GD}(x_i )=D_{\mathrm{GDA},i}^Z=D_{\mathrm{RegPA},i}^Z=D_{\mathrm{CorPA},i}^Z  \,(=\delta_i^2=\varepsilon_i)$. However, when $p\geq 3$, these methods generally do not coincide. This result triggered the research on the convergence between ORMs and GD. However, prior research comparing the ORMs and GD has been limited in scope. Some studies, such as~\cite{chao2008quantifying} and~\cite{johnson2000heuristic}, only evaluated the performance of the ORMs on restricted datasets. While~\cite{lebreton2004monte} utilized Monte Carlo simulations, their methods suffered from three notable limitations.

First, we analyzed their approach, with their computer code kindly provided by the authors, and found that it could generate correlation matrices that violated the positive (semi-)definite constraint, i.e., resulting in negative eigenvalues. This was primarily due to a procedure wherein the diagonal of the generated correlation matrix was not rescaled to unity but simply manipulated through addition. 

\begin{figure*}[t] 
\includegraphics[width=\textwidth]{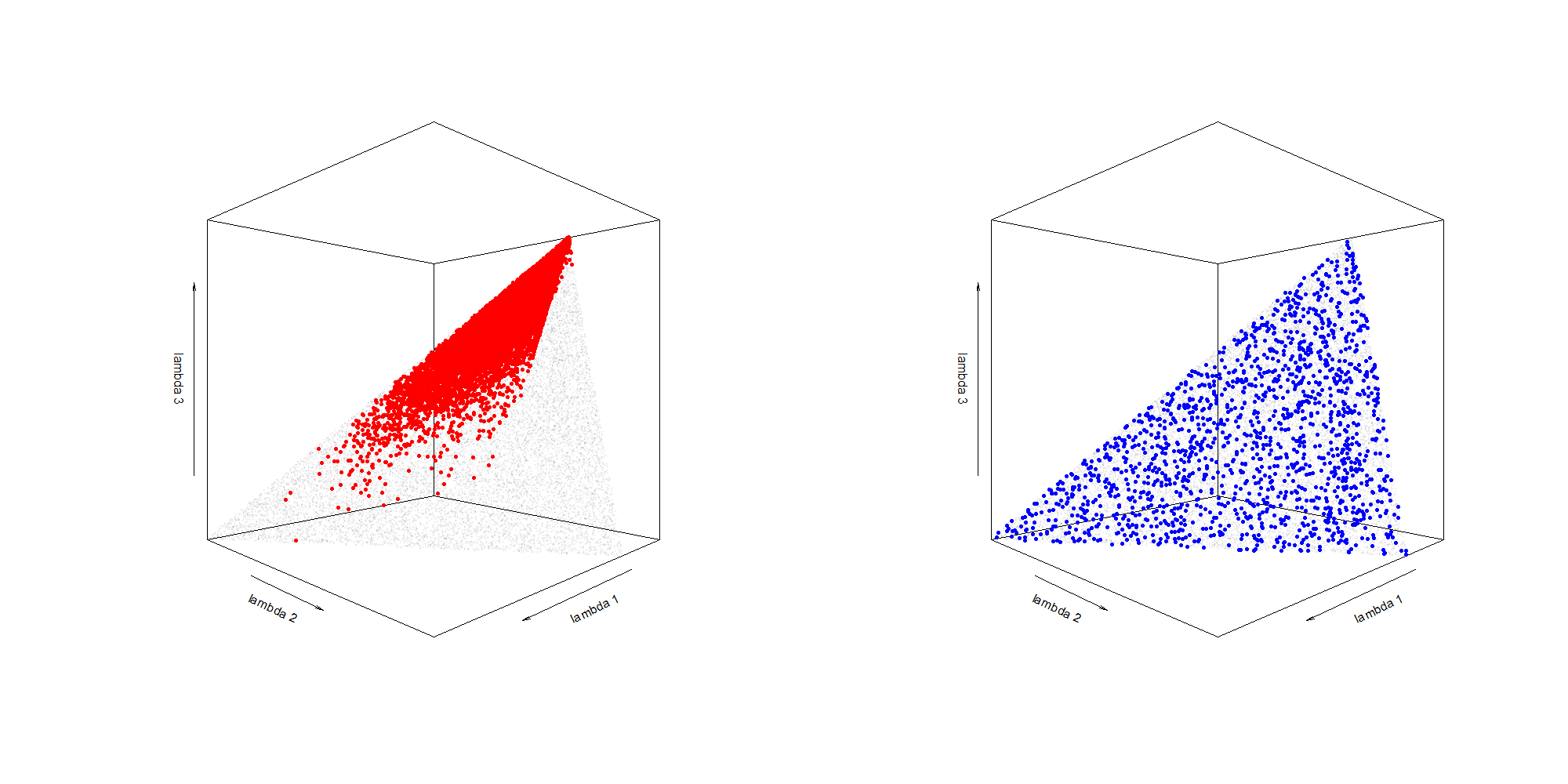}
\caption{Scatter Plots of Simulated Eigenvalues for \cite{lebreton2004monte} and Ours ($p=3$). The left panel shows the non-uniform distribution of eigenvalues produced by LeBreton et al.'s method. The right panel demonstrates the uniform distribution achieved by our approach, which shows a more even coverage across the space of all possible eigenvalues.}
\label{fig: eigenvalues of lebreton}
\end{figure*}

Secondly, recall that a predictor correlation matrix $\Sigma_{xx}$ is a positive-definite matrix given linearly independent observations of $X$. By the spectral theorem, $\Sigma_{xx}$ can be decomposed as $\Sigma_{xx}=V\Lambda V^\top$, where $\Lambda$ is a diagonal matrix with eigenvalues, and $V$ is an orthogonal matrix containing eigenvectors. Since the trace of any $p\times p$ correlation matrix is $p$, its eigenvalues must sum to $p$. Thus, all possible sets of eigenvalues lie on a $(p-1)$-dimensional simplex, denoted as $\Delta^{p-1}=\{\lambda\in\mathbb{R}^{p}:\sum_{i=1}^{p}\lambda_i=p, \lambda_1\geq\ldots\geq\lambda_p> 0\}$. A comprehensive simulation should ideally explore the full range of possible eigenvalues distributed over this entire simplex. However, we observed that the eigenvalues of correlation matrices in~\cite{lebreton2004monte} were concentrated in a specific region of this simplex. Figure~\ref{fig: eigenvalues of lebreton} visualizes the distribution of the simulated eigenvalues by the approach by LeBreton et al. and by our approach for $p=3$, with each point representing a set of eigenvalues $(\lambda_1,\lambda_2,\lambda_3)$ on $\Delta^2$. The biased eigenvalue distribution in~\cite{lebreton2004monte} may raise questions about the coverage of cases and implications of the simulation results.

\begin{figure*}[t]
\includegraphics[width=\textwidth]{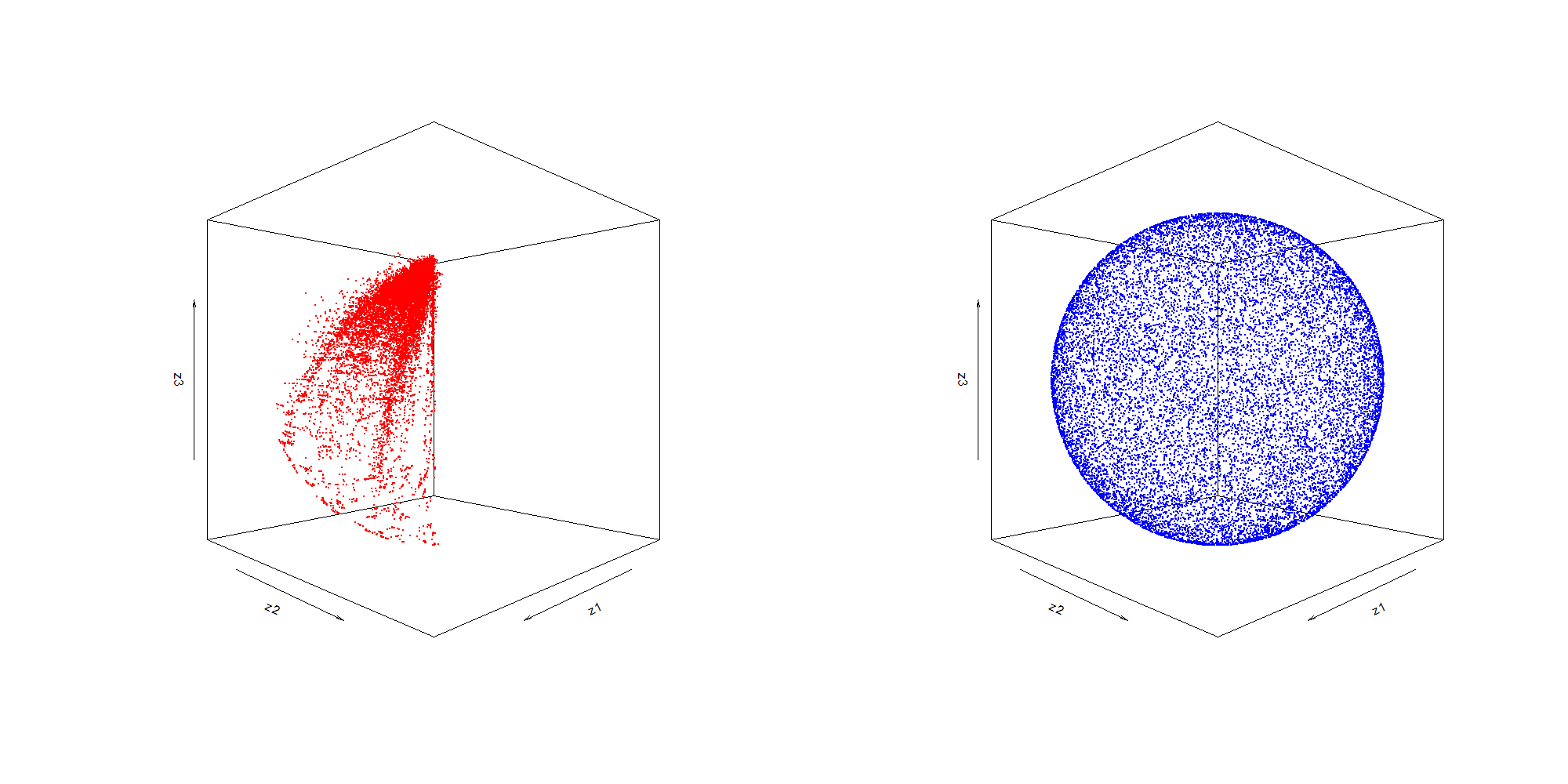}
\caption{Scatter Plots of Response Variable for \cite{lebreton2004monte} and Ours (p=3). The left panel shows the non-uniform distribution of response variables using LeBreton et al.'s approach, indicated by the clustered red points. The right panel shows our uniform distribution, where blue points are evenly spread across the unit sphere. Both plots use Johnson's Z as the orthonormal basis.}
\label{fig: responses of lebreton}
\end{figure*}

Thirdly, considering the linear model in Equation~\eqref{eq: linear regression}, it can be reformulated as
\begin{equation}
    y=Z\rho_{zy}+\epsilon.
    \label{eq: linear regression with orthogonal predictors}
\end{equation}
Here, $Z$ (Johnson’s $Z$) represents an orthonormal basis that spans the same column space as the original predictors, $X$. This reformulation allows us to conceptualize the explainable part of the response variable as a linear combination with coefficients $\rho_{zy}=Z^\top y$ due to the orthogonality of $Z$. Therefore, to construct a response variable with a specific coefficient of determination $R^2_{y\cdot X}=\sum_{i=1}^p\rho_{z_iy}^2=c$ for $c\in [0,1]$, we could start by considering a vector $u$ on the $p$-dimensional sphere of radius $\sqrt{c}$: $S_c^{p-1}=\{u\in \mathbb{R}^p:\sum_{i=1}^p u_i^2 =c\}$. This vector $u$ allows us to construct any response variable with a specified $R^2_{y\cdot X}$ in the column space of $X$. Consequently, a comprehensive simulation should ideally evaluate response variables throughout the entire $S_c^{p-1}$ However, in~\cite{lebreton2004monte}, the construction of the relationship between the response variable and predictors is very limited. As illustrated in Figure~\ref{fig: responses of lebreton}, which shows the scatter plots of the linear combination coefficients for Johnson's $Z$.~\cite{lebreton2004monte} had generated response variables covering only a portion of the sphere. Such a restricted exploration of response variables may diminish the generalizability of their findings to real-world contexts.

Moreover, the ORM can yield completely incorrect predictor importance due to the inadequately chosen orthogonalization method. Therefore, a more comprehensive simulation study that considers both reallocation and orthogonalization methods is necessary. Since J.W. Johnson’s influential paper was published in 2000, the majority of researchers have favored RW over other ORMs. Consequently, we believe comprehensive Monte Carlo simulations are warranted to thoroughly examine the performance of all ORMs under various predictor correlation structures (eigenvalues and eigenvectors) and all possible relationships with the response variables.
\section{Comprehensive Monte Carlo Simulations}
To ensure the extensiveness and coverage of cases, we explored a wide range of correlation structures for the predictors and the relationship between the response variable and predictors.
\subsection{Simulation Steps}
Our procedure for generating correlation matrices for predictors focused on manipulating both eigenvalues and eigenvectors. To evaluate all possible correlation matrices, we designed a two-step simulation procedure: (1) sampling a set of eigenvalues, and (2) generating an orthogonal matrix to construct $V\Lambda V^\top$. These steps are necessary because correlation matrices can share the same eigenvalues but differ in their eigenvectors.

For the first step, recall that all possible sets of eigenvalues lie on a $(p-1)$-dimensional simplex, $\Delta^{p-1}$. To ensure broad coverage, we uniformly sampled eigenvalue sets from $\Delta^{p-1}$. This is done by Kraemer’s Algorithm~\citep{smith2004sampling}. The second step, generating orthogonal matrices, presented a challenge. While there are methods to sample orthogonal matrices uniformly, these alone cannot guarantee that $V\Lambda V^\top$ will result in a valid correlation matrix without further adjustments.

Consequently, we relied on the Method of Alternating Projection (MAP;~\cite{waller2020generating}) in our Monte Carlo simulations to generate correlation matrices based on sets of uniformly sampled eigenvalues. To introduce variability, we used random seeds embedded in the MAP algorithm to generate different orthogonal matrices for each sampled eigenvalue set. Specifically, we run MAP ten times to produce correlation matrices that share the same eigenvalues but have different eigenvectors. This variation is crucial, as the eigenvectors directly influence the Variance Inflation Factor (VIF) of the predictors. We implemented the MAP procedure using the \texttt{rMAP} from \texttt{R} package \texttt{fungible}.

Unlike~\cite{lebreton2004monte}, who directly generated a correlation matrix for both predictors and response, we decouple the generation process, first creating a correlation structure for the predictors and then, independently, generating response variables for a given predictor correlation structure. We use Equation~\eqref{eq: expected performance metric}
\begin{equation}
    \mathbb{E}_y[f(\mathrm{GD}(X), D_{A}^{\tilde{Z}})|X],
    \label{eq: expected performance metric}
\end{equation}
where $f$ denotes performance metric, \[\mathrm{GD}(X)=[\mathrm{GD}(x_1),\ldots,\mathrm{GD}(x_p)]^\top\] and $D_A^{\tilde{Z}}=[D_{A,1}^{\tilde{Z}},\ldots,D_{A,p}^{\tilde{Z}}]^\top$, to evaluate the performance of ORMs. Conceptually, Equation~\eqref{eq: expected performance metric} considers the expected performance of ORMs across all possible response variables. Given this objective, we can generate all possible response variables by uniformly sampling vectors from the $p$-dimensional unit sphere. The specific details of how we generate such response variables are provided in Appendix~\ref{app: A}.

We believe that the utility of the ORMs lies in their ability to approximate the importance of predictors evaluated by GD. We use the root mean square error (RMSE) of the ORMs with respect to GD to measure the quantitative differences between them. In addition, as~\cite{lebreton2004monte} did in a prior Monte Carlo simulation, we use Kendall's $\tau$~\citep{kendall1938new} to evaluate the ordinal association between GD and the ORMs. The general test procedures are stated as follows.
\begin{enumerate}
    \item Varying the number of predictors ($p$).\\
    We considered eight levels of $p:p\in\{3,\ldots,10\}$. For each $p$, subsequent steps were repeated.

    \item 	Sampling eigenvalues ($\lambda$).\\
    For each $k\in\{1,2,\ldots,n_{\mathrm{ev}}\}$:
    \begin{itemize}
        \item Uniformly sample a vector $\lambda$ from the $(p-1)$-dimensional simplex, $\Delta^{p-1}=\{\lambda\in\mathbb{R}^p:\sum_{i=1}^p \lambda_i=p, \lambda_1\geq \ldots\geq \lambda_p>0\}$.
    \end{itemize} 

    \item 	Generating predictor correlation matrices.\\ 
    For each seed $j\in\{1,2,\ldots,n_s\}$:
    \begin{itemize}
        \item Generate a correlation matrix for predictors, $\Sigma_{xx}$, by applying the function \[\texttt{rMAP}(\text{eigenvalues}=\lambda, \text{seed}=j).\] This procedure ensures that $\Sigma_{xx}$ is a valid, positive-definite correlation matrix determined by the sampled eigenvalues and appropriate eigenvectors.
    \end{itemize}

    \item Sampling the coefficients of response vector ($u$). \\
    For each $i\in\{1,2,\ldots,n_u\}$:
    \begin{itemize}
        \item Uniformly sample a vector $u$ from the $p$-dimensional unit sphere,
        \[S^{p-1}=\left\{u\in\mathbb{R}^p:\sum_{i=1}^p u_i^2=1\right\}.\]
        \item Compute the correlation between the predictors and the response variable, $\rho_{xy}$. This results in a $(p+1)\times (p+1)$ correlation matrix, $\Sigma_{yx}$, that includes both $\Sigma_{xx}$ and the correlations with the response.
    \end{itemize}

    \item Computing relative importance measures.\\
    Using the augmented correlation matrix $\Sigma_{yx}$, compute the relative importance measures: $\mathrm{GD}(x_i)$ and $D_{A,i}^{\tilde{Z}}$ where $A\in\{\mathrm{GDA},\mathrm{CorPA},\mathrm{RegPA},\mathrm{IdA}\}$ and $\tilde{Z}\in\{Z,Z_{\mathrm{GS}},Z_{\mathrm{PC}},Z_{\mathrm{VM}}\}$ for each predictor.

    \item Evaluating performance metrics.
    \begin{itemize}
        \item Root mean squared error (RMSE).
        \item Kendall's $\tau$.
    \end{itemize}
    
\end{enumerate}
We set the number of the set of eigenvalues $n_{\mathrm{ev}}=1000$ for relatively small number of predictors: $p\in\{3,4,5,6\},$ and $n_{\mathrm{ev}}=2500$ for relatively large number of predictors: $p\in\{7,8,9,10\}$ to ensure broad coverage. The number of the seed $n_s=10$, and the number of coefficients of response vector $n_u=100$. The tests were implemented using the \texttt{R} statistical programming environment~\citep{r2024r}. Code to replicate the simulation and empirical examples are available at \url{https://github.com/tien-endotchang/ORMs}.
\subsection{Results on the Orthogonalization}
We first examine the test results on the orthogonalization methods. We averaged the performance metrics across the response variables, the eigenvectors and the eigenvalues and summarized them in Table~\ref{tab: mean result for z and a}. The results show that, across all reallocation methods, Johnson's $Z$ consistently outperforms other orthogonalization methods in both Kendall's $\tau$ and RMSE. In contrast, the Gram-Schmidt orthogonalization ($Z_{\mathrm{GS}}$) and principal component ($Z_{\mathrm{PC}}$) methods perform quite poorly in some cases. Unexpectedly, while the performance of the Varimax ($Z_{\mathrm{VM}}$) falls short of Johnson's $Z$, it is the closest to Johnson’s $Z$ among all other orthogonalization methods we examined.

The superiority of Johnson’s $Z$ may stem from its definition: the orthogonalization that requires the minimal adjustment in the least-squares sense. This definition is equivalent to maximizing the simple correlation between each original predictor $x_i$ and its corresponding orthogonal transformation $z_i$. As a result, the importance of an original predictor $x_i$ should be primarily accounted for by $z_i$. The relatively better performance of Varimax $Z_{\mathrm{VM}}$ may be attributed to some shared geometric properties with Johnson's $Z$. The objective function in Equation~\eqref{eq: varimax} can be seen as minimizing the weighted complexity between rows and columns of a loading matrix, which is used to achieve a simple structure in a loading matrix~\citep{browne2001overview}. We found that the simple structure seems to have a connection to the minimal transformation. We conjecture that Johnson’s $Z$ and $Z_{\mathrm{VM}}$ are identical in the two predictors case (a preliminary proof is attached in Appendix~\ref{app: B}). In contrast, the poor performance of $Z_{\mathrm{GS}}$ and $Z_{\mathrm{PC}}$ may result from their inability to provide a simple structure in a loading matrix.

\begin{table*}[tb]
\caption{Mean RMSE/Kendall's tau across four orthogonalization methods and four reallocation methods.}
\label{tab: mean result for z and a}
\centering
\small
\setlength{\tabcolsep}{3pt}
\begin{tabular}{cccccccccc}
\hline
$A$ & $\tilde{Z}$  & $p = 3$        & $p = 4$        & $p = 5$        & $p = 6$        & $p = 7$        & $p = 8$        & $p = 9$       & $p = 10$      \\ \hline
\multirow{4}{*}{GDA} &
  \textbf{Johnson} &
  \textbf{0.01 / 0.94} &
  \textbf{0.01 / 0.93} &
  \textbf{0.01 / 0.92} &
  \textbf{0.01 / 0.92} &
  \textbf{0.01 / 0.91} &
  \textbf{0.01 / 0.90} &
  \textbf{0.01 / 0.90} &
  \textbf{0.01 / 0.90} \\
  & GS & 0.08 / 0.62  & 0.07 / 0.63  & 0.06 / 0.65  & 0.06 / 0.64  & 0.05 / 0.63  & 0.05 / 0.63  & 0.04 / 0.62 & 0.04 / 0.62 \\
  & PC & 0.15 / 0.27  & 0.12 / 0.36  & 0.10 / 0.35   & 0.09 / 0.36  & 0.08 / 0.34  & 0.08 / 0.34  & 0.07 / 0.32 & 0.07 / 0.30 \\
  & VM & 0.02 / 0.89  & 0.02 / 0.86  & 0.02 / 0.85  & 0.02 / 0.84  & 0.02 / 0.83  & 0.02 / 0.82  & 0.02 / 0.81 & 0.02 / 0.81 \\ \hline
\multirow{4}{*}{CorPA} &
  \textbf{Johnson} &
  \textbf{0.02 / 0.94} &
  \textbf{0.02 / 0.93} &
  \textbf{0.02 / 0.91} &
  \textbf{0.01 / 0.91} &
  \textbf{0.01 / 0.90} &
  \textbf{0.01 / 0.89} &
  \textbf{0.01 / 0.89} &
  \textbf{0.01 / 0.88} \\ 
  & GS & 0.08 / 0.74 & 0.07 / 0.71 & 0.06 / 0.70 & 0.05 / 0.67 & 0.05 / 0.68 & 0.04 / 0.66 & 0.04 / 0.67 & 0.03 / 0.66 \\
  & PC & 0.20 / 0.01 & 0.17 / 0.00 & 0.15 / -0.05 & 0.14 / -0.05 & 0.12 / -0.03 & 0.10 / 0.01 & 0.09 / 0.02 & 0.08 / 0.00 \\
  & VM & 0.02 / 0.90 & 0.03 / 0.88 & 0.03 / 0.86 & 0.03 / 0.85 & 0.02 / 0.84 & 0.02 / 0.83 & 0.02 / 0.82 & 0.02 / 0.82 \\ \hline
\multirow{4}{*}{RegPA} &
  \textbf{Johnson} &
  \textbf{0.02 / 0.92} &
  \textbf{0.02 / 0.89} &
  \textbf{0.02 / 0.88} &
  \textbf{0.02 / 0.86} &
  \textbf{0.02 / 0.85} &
  \textbf{0.02 / 0.84} &
  \textbf{0.02 / 0.83} &
  \textbf{0.02 / 0.84} \\ 
  & GS & 0.08 / 0.73 & 0.07 / 0.71 & 0.06 / 0.68 & 0.06 / 0.67 & 0.05 / 0.66 & 0.05 / 0.64 & 0.04 / 0.65 & 0.04 / 0.64 \\
  & PC & 0.20 / 0.01 & 0.17 / 0.00 & 0.15 / -0.05 & 0.14 / -0.05 & 0.12 / -0.03 & 0.10 / 0.01 & 0.09 / 0.02 & 0.08 / 0.00 \\
  & VM & 0.02 / 0.91 & 0.02 / 0.88 & 0.02 / 0.87 & 0.02 / 0.85 & 0.02 / 0.84 & 0.02 / 0.83 & 0.02 / 0.82 & 0.02 / 0.82 \\ \hline
\multirow{4}{*}{IdA} &
  \textbf{Johnson} &
  \textbf{0.05 / 0.90} &
  \textbf{0.05 / 0.87} &
  \textbf{0.04 / 0.84} &
  \textbf{0.04 / 0.83} &
  \textbf{0.03 / 0.83} &
  \textbf{0.03 / 0.82} &
  \textbf{0.03 / 0.82} &
  \textbf{0.03 / 0.81} \\
  & GS & 0.12 / 0.58 & 0.10 / 0.57 & 0.09 / 0.58 & 0.08 / 0.56 & 0.07 / 0.55 & 0.07 / 0.54 & 0.06 / 0.54 & 0.05 / 0.54 \\
  & PC & 0.29 / -0.05 & 0.24 / -0.02 & 0.21 / -0.02 & 0.19 / -0.03 & 0.16 / -0.01 & 0.15 / -0.01 & 0.13 / 0.01 & 0.12 / 0.01 \\
  & VM & 0.06 / 0.85 & 0.06 / 0.80 & 0.05 / 0.76 & 0.05 / 0.75 & 0.04 / 0.74 & 0.04 / 0.73 & 0.04 / 0.72 & 0.04 / 0.72 \\ \hline
\multicolumn{10}{p{0.95\textwidth}}{\small Note: The numbers represent RMSE / Kendall’s $\tau$, respectively. $A$: reallocation method; $\tilde{Z}$: orthogonalization method; $p$: number of predictors.} \\
\end{tabular}
\end{table*}
\subsubsection{\texorpdfstring{An optimal $\tilde{Z}^\star$ better than Johnson's $Z$?}{An optimal Z~* better than Johnson's Z?}}
Our comprehensive Monte Carlo simulations reveal that Johnson's $Z$ demonstrates superior performance among common orthogonalization methods. This raises the question: Is there an optimal $\tilde{Z}^{\star}$ that could outperform Johnson’s $Z$? Here, we offer another possibility to find the best orthogonalization. One simple idea is to find the $\tilde{Z}^{\star}$ that minimizes the expected RMSE across all possible response variables $y$. Take $D_{\mathrm{CorPA}}^{\tilde{Z}}$ for example. If we aim to find an orthogonalization method that minimize the RMSE between GD and $D_{\mathrm{CorPA}}^{\tilde{Z}}$, then, based on Equation~\eqref{eq: expected performance metric}, the optimal $\tilde{Z}^{\star}$ can be obtained by minimizing 
\begin{equation}
    \min_{\tilde{Z}} \mathbb{E}_y \left[\sqrt{\sum_{i=1}^p \left(\mathrm{GD}(x_i)-D_{\mathrm{CorPA}}^{\tilde{Z}}\right)^2}|X\right].
    \label{eq: obj for optimal z}
\end{equation}
While this formulation appears complex, it becomes tractable when we incorporate $\tilde{Z}$ into the calculation of GD for $p=3$. Moreover, the expectation over $y$ is readily computable, as we assume that the explainable part of $y$ is uniformly distributed within the space spanned by $\tilde{Z}$ (see Appendix~\ref{app: A} for details). We then solve this minimization problem by Wolfram Mathematica~\citep{Mathematica2025}. The resulting $D_{\mathrm{CorPA}}^{\tilde{Z}^\star}$ does demonstrate superior performance compared to Johnson’s $Z$. Not unexpectedly, we found that Johnson's $Z$ is very close to the optimal $\tilde{Z}^\star$ for CorPA when measured by Euclidean distance.

These findings indicate the crucial impact of orthogonalization methods on the performance of ORMs. While there is a possibility for other orthogonalization methods outperforming Johnson's $Z$, we have found its performance to be generally robust and competitive. Hence, we will use Johnson's $Z$ as the orthogonalization method in the analyses that follow.
\subsection{Results on the Reallocation}
For each correlation structure, we averaged the performance metrics across the response variables. Recall that $D_{\mathrm{IdA},i}^Z=w_i^2; D_{\mathrm{RegPA},i}^Z=\delta_i^2; D_{\mathrm{CorPA},i}^Z=\varepsilon_i$ where the last two are referred to as GCD and RW and we should use these terms interchangeably. Through our comprehensive Monte Carlo simulations, we found that the test results on the reallocation methods can be categorized into four scenarios based on two key factors: $\lambda_1/\sqrt{p}$ and $\mathrm{VIF}_{\max}/p$ where $\lambda_1$ denotes the largest eigenvalue of the correlation matrix $\Sigma_{xx}=X^\top X$ and the variance inflation factor (VIF) for predictor $x_i$, a measure of multicollinearity, is calculated as:
\begin{equation}
    \mathrm{VIF}_i=\frac{1}{1-R^2_{x_i\cdot X_{-i}}},
    \label{eq: vif}
\end{equation}
where $X_{-i}$ is the matrix of all predictors except $x_i$. We define $\mathrm{VIF}_{\max}$ as the largest $\mathrm{VIF}_i$ across all predictors $(i=1,\ldots,p)$. With extensive exploration of a wide range of eigenvalue and VIF values, we have found that $\lambda_1/\sqrt{p}=1.5$ and $\mathrm{VIF}_{\max}/p=4$ can be used as criteria to define four scenarios for practical guidelines on how to use the ORMs. The scenarios are outlined in Table~\ref{tab: four scenarios}. 

\begin{table}[htb]
\centering
\caption{The summarized scenarios based on $\lambda_1/\sqrt{p}$ and $\mathrm{VIF}_{\max}/p$.}
\label{tab: four scenarios}
\begin{tabular}{ccc}
\hline
       & $\frac{\lambda_1}{\sqrt{p}}<1.5$     & $\frac{\lambda_1}{\sqrt{p}}\geq 1.5$    \\ \hline
$\frac{\mathrm{VIF}_{\max}}{p}<4$     & Scenario 1.1     & Scenario 1.2    \\
$\frac{\mathrm{VIF}_{\max}}{p}\geq 4$     & Scenario 2.1     & Scenario 2.2    \\ \hline
\end{tabular}
\end{table}
Since $\lambda_1$ represents the magnitude of first principal component, and $\mathrm{VIF}_{\max}$ indicates the severity of multicollinearity among predictors, we name the four scenarios as mild/severe multicollinearity with mild/strong first principal component.
\subsubsection{\texorpdfstring{Scenario 1.1: Mild Multicollinearity $(\mathrm{VIF}_{\max}/p<4)$ with Mild First Principal Component $(\lambda_1/\sqrt{p}<1.5)$}{Scenario 1.1: Mild Multicollinearity (VIFmax/p<4) with Mild First Principal Component (lambda1/sqrt(p)<1.5)}}

\begin{figure*}[tb]
\includegraphics[width=\textwidth]{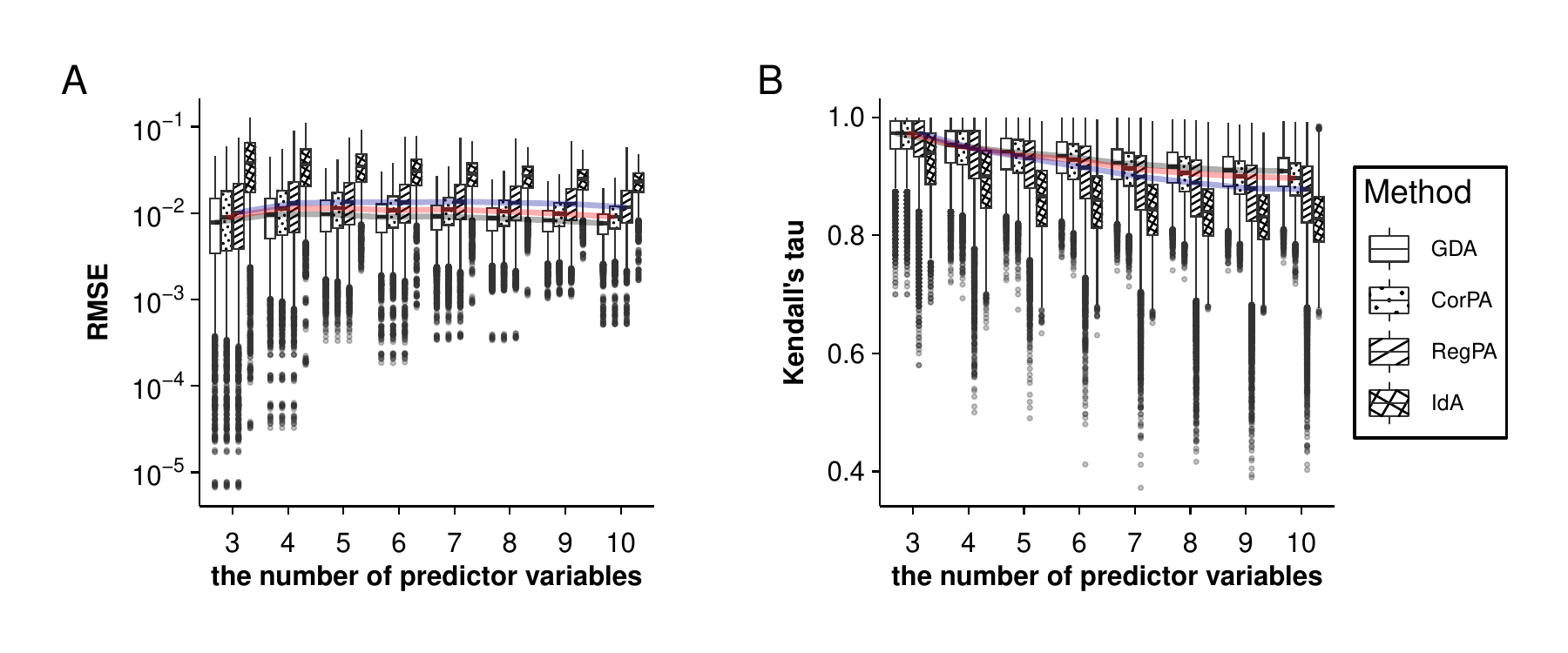}
\caption{Scenario 1.1 performance comparison of four reallocation methods using Johnson's Z in ORMs, assessed by RMSE (left) and Kendall's tau (right). Red line indicates CorPA (RW)'s median performance, while blue and gray lines show RegPA (GCD) and GDA results, respectively.}
\label{fig: s1.1}
\end{figure*}

In Scenario 1.1 (Figure~\ref{fig: s1.1}), all four reallocation methods exhibit comparable performance. As a benchmark, $\mathrm{GDA}^Z$ demonstrates the lowest RMSE (Figure~\ref{fig: s1.1}A) and highest Kendall's tau (Figure~\ref{fig: s1.1}B). The methods rank similarly in both metrics: $\mathrm{GDA}^Z$, followed closely by $\mathrm{CorPA}^Z$, then $\mathrm{RegPA}^Z$, and lastly, $\mathrm{IdA}^Z$. The median performance of $\mathrm{GDA}^Z$, $\mathrm{CorPA}^Z$ and $\mathrm{RegPA}^Z$, in gray, red and blue lines, respectively, are also shown in Figure~\ref{fig: s1.1}. These results suggest that $\mathrm{CorPA}^Z$ proves to be a favorite choice under mild multicollinearity and mild first principal component.

\subsubsection{\texorpdfstring{Scenario 1.2: Mild Multicollinearity $(\mathrm{VIF}_{\max}/p<4)$ with Strong First Principal Component $(\lambda_1/\sqrt{p}\geq 1.5)$}{Scenario 1.2: Mild Multicollinearity (VIFmax/p<4) with Strong First Principal Component (lambda1/sqrt(p)>=1.5)}}

\begin{figure*}[tb]
\includegraphics[width=\textwidth]{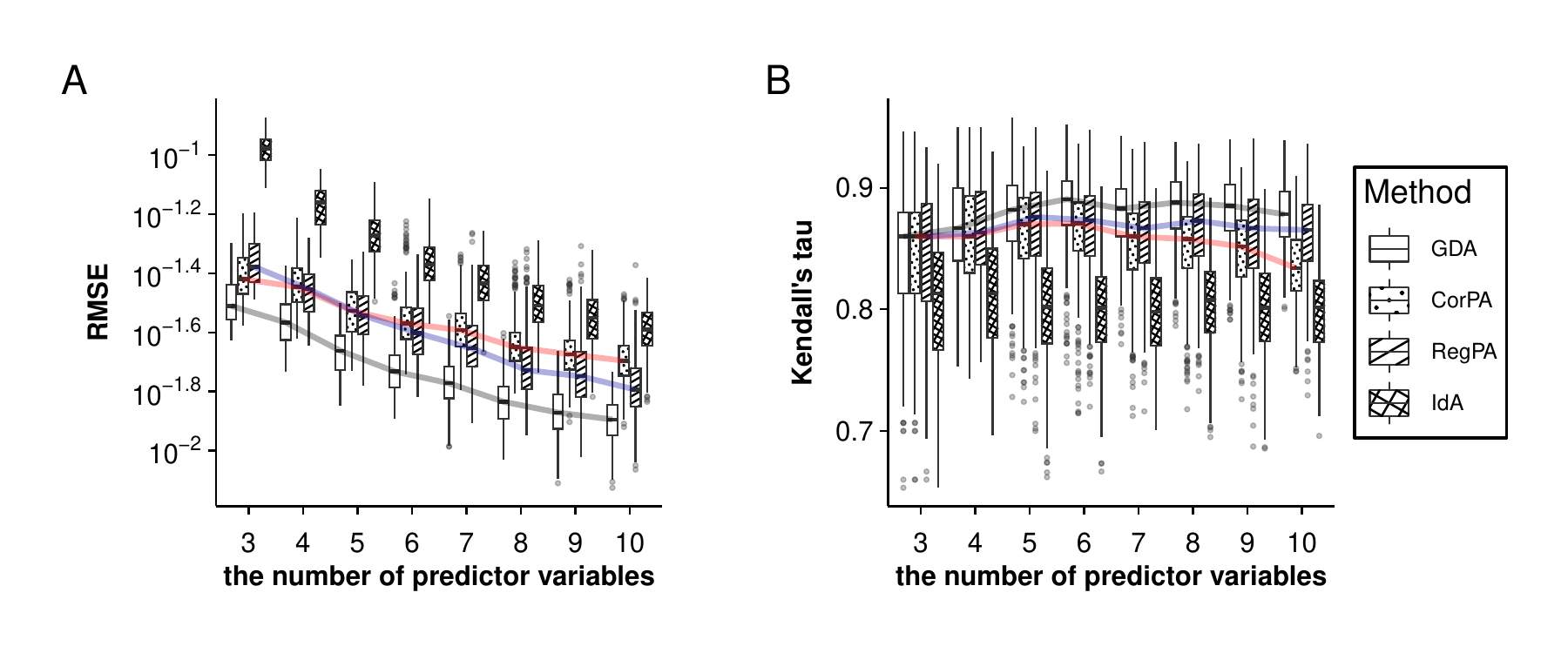}
\caption{Scenario 1.2 performance comparison of four reallocation methods using Johnson's Z in ORMs, assessed by RMSE (left) and Kendall's tau (right). Red line indicates CorPA (RW)'s median performance, while blue and gray lines show RegPA (GCD) and GDA results, respectively.}
\label{fig: s1.2}
\end{figure*}

In Scenario 1.2 (Figure~\ref{fig: s1.2}), $\mathrm{GDA}^Z$, as a benchmark, again demonstrates the best performance in terms of both RMSE (Figure~\ref{fig: s1.2}A) and Kendall’s tau (Figure~\ref{fig: s1.2}B). Notably, $\mathrm{RegPA}^Z$ outperforms $\mathrm{CorPA}^Z$ in both RMSE and Kendall’s tau as the number of predictors increases. The results suggest that RW is not always favorable, especially when the first principal component is strong.

\begin{figure*}[tb]
\includegraphics[width=\textwidth]{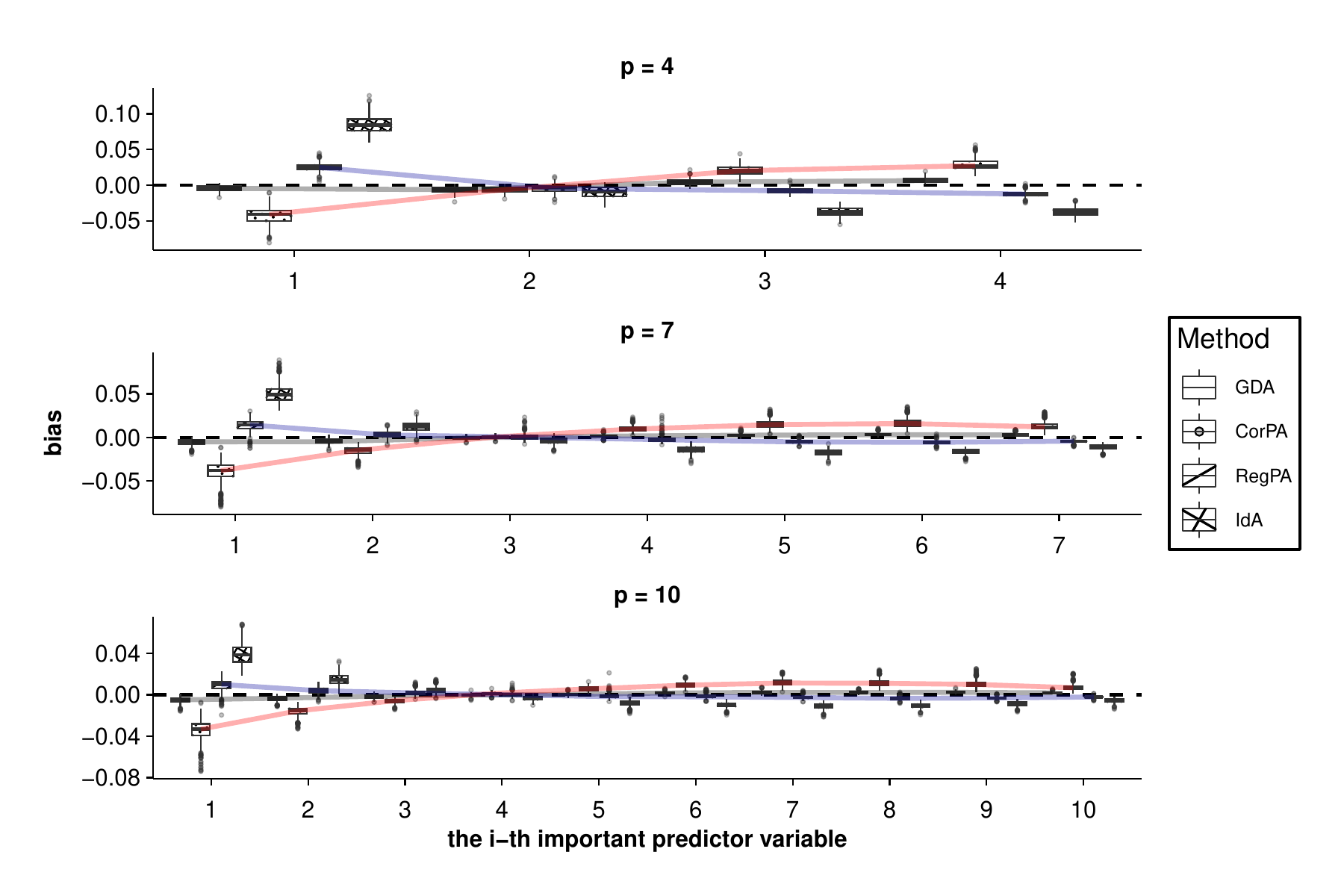}
\caption{Demonstration of the leveling problem of RW in Scenario 1.2. The x-axis represents the $i$-th important predictor determined by GD. Red line indicates CorPA (RW)'s median performance, while blue and gray lines show RegPA (GCD) and GDA results, respectively.}
\label{fig: leveling}
\end{figure*}

This phenomenon can be explained by \emph{the leveling problem} of RW. Recall that $D_{\mathrm{CorPA}}^Z=\varepsilon_i=\sum_{j=1}^p \ell_{Z,ij}^2 \rho_{z_j y}^2$ where $\ell_{Z,ij}$ is equivalent to the correlation between $z_j$ and $x_i$. Note that $\sum_{j=1}^p \ell_{Z,ij}^2=1,\forall i=1,\ldots,p$. Thus, $\varepsilon_i$ can be interpreted as a weighted average of $\rho_{z_j y}^2$ , with weights given by $\ell_{Z,ij}^2$. Ideally, these weights should vary based on the relationships between each $z_j$ and $x_i$, but when a strong first principal component exists (i.e., when $\lambda_1/\sqrt{p}$ is substantially large), the weights $\ell_{Z,ij}^2$ tend to become similar for all $j$. This occurs because the predictors $x_i$ become highly correlated and point in similar directions in the vector space, with small angles between them. While Johnson's $Z$ aims to minimize the difference between $x_i$ and $z_j$, the $z_j$ variables are constrained to be orthogonal. As a result, most $x_i$ variables will have a larger angle with $z_j$, leading to small $\ell_{Z,ij}^2$ for most $i,j$. Consequently, $\ell_{Z,ij}^2$ approaches $1/p$ as $\lambda_1/\sqrt{p}$ becomes large. We call this the leveling problem because it leads to a low trace of the reallocation matrix and pulls the importance of all predictors towards an average. Specifically, Figure~\ref{fig: leveling} shows the estimation bias, $D_{A,i}^Z-\mathrm{GD}(x_i)$, by ORMs, the x-axis indicates the $i$-th important predictor determined by GD. It can be clearly seen that RW will underestimate the importance of stronger predictors and overestimate that of weaker ones, causing RW's performance to deteriorate in both Kendall's tau and RMSE. In contrast, the bias of GDA and GCD does not exhibit this trend.

\begin{figure*}[tb]
\includegraphics[width=\textwidth]{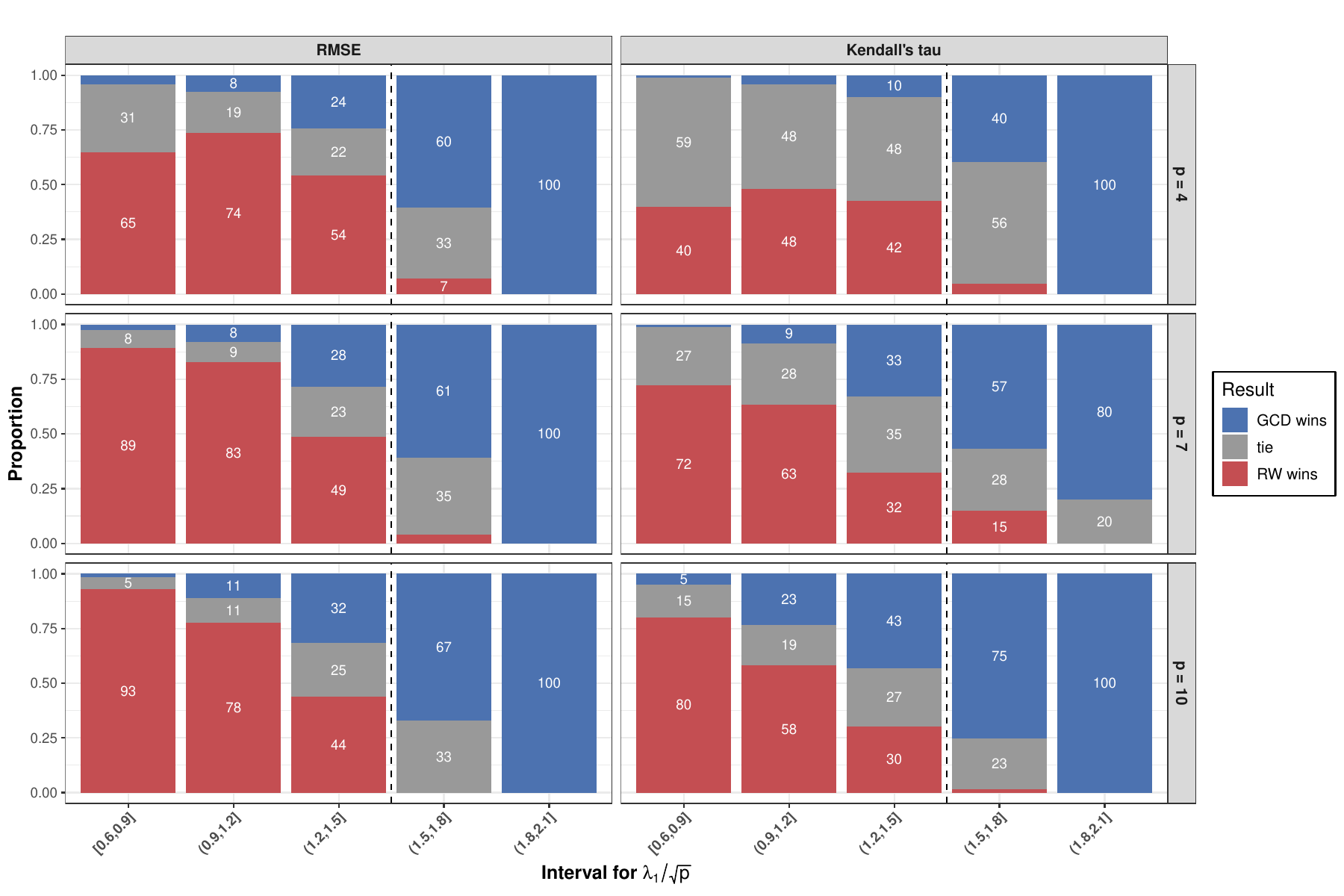}
\caption{Win-Loss Analysis between RW and GCD with mild multicollinearity (Scenario 1.1 and 1.2) in RMSE and Kendall’s tau. The dashed line indicates the threshold value $\lambda_1/\sqrt{p}=1.5$.}
\label{fig: s1.1 and 1.2 win-loss}
\end{figure*}

The consequences of the RW leveling problem are demonstrated through a win-loss analysis. Using our simulation framework, each eigenvalue set was paired with ten correlation matrices with varied eigenvectors. We used Wilcoxon signed-rank pair-wise tests to compare RW and GCD performance across these ten replicates for each set of eigenvalues, classifying each result as a RW win, a GCD win, or a tie. As demonstrated in Figure~\ref{fig: s1.1 and 1.2 win-loss}, in mild multicollinearity conditions (Scenarios 1.1 and 1.2), RW's performance deteriorates with increasing $\lambda_1/\sqrt{p}$. The number of wins for RW decreases, confirming the growing severity of its leveling problem. The dashed line in Figure~\ref{fig: s1.1 and 1.2 win-loss} shows the threshold value at $\lambda_1/\sqrt{p}=1.5$, which marks a switch from RW outperforming GCD to the converse. This analysis underscores the need to account for predictor correlation structure for selecting an appropriate ORM. 

\subsubsection{\texorpdfstring{Scenario 2.1: Severe Multicollinearity $(\mathrm{VIF}_{\max}/p\geq 4)$ with Mild First Principal Component $(\lambda_1/\sqrt{p}<1.5)$}{Scenario 2.1: Severe Multicollinearity (VIFmax/p>=4) with Mild First Principal Component (lambda1/sqrt(p)<1.5)}}

\begin{figure*}[tb]
\includegraphics[width=\textwidth]{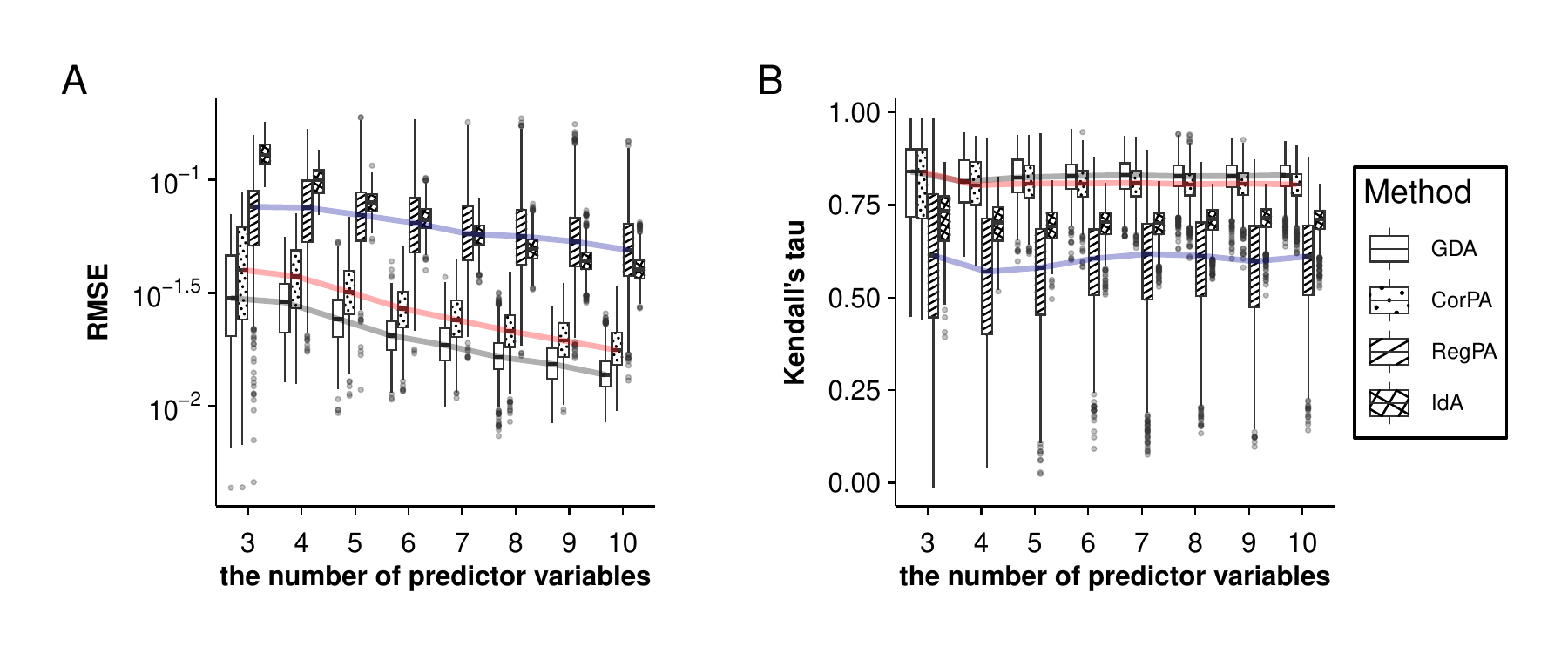}
\caption{Scenario 2.1 performance comparison of four reallocation methods using Johnson's $Z$ in ORMs, assessed by RMSE (left) and Kendall's tau (right). Red line indicates CorPA (RW)'s median performance, while blue and gray lines show RegPA (GCD) and GDA results, respectively.}
\label{fig: s2.1}
\end{figure*}

In Scenario 2.1 (Figure~\ref{fig: s2.1}), $\mathrm{RegPA}^Z$'s performance in terms of both RMSE and Kendall’s tau declines notably, as shown in Figure~\ref{fig: s2.1}A and~\ref{fig: s2.1}B.

This can be explained by \emph{the a priori bias} of GCD. Considering a uniformly distributed random variable $u$ on a unit sphere $S^{p-1}$. As we introduced in previous section, we can construct the response variable uniformly with $u$ as the coefficients and $Z$ as the basis. By the symmetry of this uniform distribution, we have $\mathbb{E}[u_i]=0,\forall i=1,\ldots,p$ and $\mathbb{E}[u_i u_j]=0,\forall i\neq j$. It can be also easily shown $\mathbb{E}[u_i^2]=1/p,\forall i=1,\ldots,p$. Based on these properties, we can derive the expected values of GD to be (see Appendix~\ref{app: B} for proofs)
\begin{equation}
    \mathbb{E}_y[\mathrm{GD}(x_i)]=\frac{R^2_{y\cdot X}}{p},\quad \forall i=1,\ldots,p
    \label{eq: expected GD}
\end{equation}
and the expected values of the ORMs with Johnson’s $Z$ and an arbitrary reallocation to be
\begin{equation}
    \mathbb{E}_y[D_{A,i}^{Z}]=\frac{R^2_{y\cdot X}}{p}\sum_{j=1}^p a_{ij}^Z,\quad \forall i=1,\ldots,p.
    \label{eq: expected ORM}
\end{equation}
Comparing Equation~\eqref{eq: expected GD} to~\eqref{eq: expected ORM}, we may expect that if the ORM is to approximate the GD, then the two expectations should be the same. That is, the row-sum of the reallocation matrix $\sum_{j=1}^p a_{ij}^Z$ should be one. In addition, Equation~\eqref{eq: expected ORM} implies the row-sum has an a priori effect on the output of the ORM. Specifically, if the $i$-th row-sum is significantly greater or smaller than one, then the ORM for the predictor $x_i$ will tend to be overestimated or underestimated, respectively, regardless of the response variable $y$. Such an estimate bias is determined a priori by the row-sum of the reallocation matrix. That is, the row-sum somewhat predetermines the importance of a predictor even without knowing what the $y$ is. In contrast, when Johnson's $Z$ is used, it can be easily shown that $\mathrm{GDA}^Z, \mathrm{IdA}^Z$ and $\mathrm{CorPA}^Z$ all have $\sum_{j=1}^p a_{ij}^Z=1,\forall i=1,\ldots,p$ (for GDA, the proof is in the Appendix~\ref{app: B}). Only $\mathrm{RegPA}^Z$ does not have this property in general, which explains the high variation and poor performance of GCD in this scenario.

We illustrate the mechanism behind the a priori bias of GCD. Let VIF a $p\times 1$ vector whose elements are $\mathrm{VIF}_i,\forall i=1,\ldots,p$. Then we can write VIF as
\begin{equation}
    \mathrm{VIF}=\mathrm{diag}^{-1}[(X^\top X)^{-1}]=\mathrm{diag}^{-1}[\Gamma_Z^\top\Gamma_Z],
    \label{eq: VIF matrix form}
\end{equation}
where $\mathrm{diag}^{-1}[\cdot]$ extracts the diagonal elements of a matrix. This also indicates that $\sum_{i=1}^p \gamma_{Z,ij}^2=\mathrm{VIF}_j$. Combining this with Equation~\eqref{eq: RegPA with minimal trans}, we have
\begin{equation}
    \mathrm{RegPA}_{ij}^Z=\gamma_{ij}^{*2}=\frac{\gamma_{Z,ij}^2}{\mathrm{VIF}_j}.
    \label{eq: RegPA and VIF}
\end{equation}
Although the $\mathrm{RegPA}^Z$ in Equation~\eqref{eq: RegPA and VIF} fulfills the properties mentioned in the section of Reallocation, it results in severe bias problem when there is a dominant $\mathrm{VIF}_{\max}$. Let $k=\arg\max_i\mathrm{VIF}_i$. Because $\Gamma_Z$ is symmetric, we have $\sum_{j=1}^p\gamma_{Z,kj}^2=\sum_{i=1}^p \gamma_{Z,ik}^2=\mathrm{VIF}_k$. The $k$-th row-sum of $\mathrm{RegPA}^Z$ is defined as
\begin{align}
    \sum_{j=1}^p\mathrm{RegPA}_{kj}^Z &= \sum_{j=1}^p\frac{\gamma_{Z,kj}^2}{\mathrm{VIF}_j} \notag\\
    &=\frac{\gamma_{Z,k1}^2}{\mathrm{VIF}_1}+\frac{\gamma_{Z,k2}^2}{\mathrm{VIF}_2}+\ldots+\frac{\gamma_{Z,kp}^2}{\mathrm{VIF}_p} \notag \\
    &>\frac{\gamma_{Z,k1}^2}{\mathrm{VIF}_k}+\frac{\gamma_{Z,k2}^2}{\mathrm{VIF}_k}+\ldots+\frac{\gamma_{Z,kp}^2}{\mathrm{VIF}_k} \notag \\
    &=\frac{\sum_{j=1}^p \gamma_{Z,kj}^2}{\mathrm{VIF}_k} \notag \\
    &=\frac{\mathrm{VIF}_k}{\mathrm{VIF}_k}=1.
    \label{eq: row-sum and VIF}
\end{align}
Equation~\eqref{eq: row-sum and VIF} implies that the $k$-th row-sum is always greater than one. Since $\mathrm{VIF}_k=\mathrm{VIF}_{\max}$ is significantly larger than other VIF’s, the $k$-th row-sum will also be significantly larger than one, leading to an a priori bias of overestimation. Furthermore, since the total sum of $\mathrm{RegPA}_{ij}^Z$ is a constant $p$, when some row-sums are larger than one, other row-sums have to be smaller than one, leading to a priori bias of underestimation, causing GCD's performance to deteriorate in both RMSE and Kendall's tau. In this scenario, $\mathrm{CorPA}^Z$ again proves to be a favorable ORM.
\subsubsection{\texorpdfstring{Scenario 2.2: Severe Multicollinearity $(\mathrm{VIF}_{\max}/p\geq 4)$ with Strong First Principal Component $(\lambda_1/\sqrt{p}\geq 1.5)$}{Scenario 2.2: Severe Multicollinearity (VIFmax/p>=4) with Strong First Principal Component (lambda1/sqrt(p)>=1.5)}}

\begin{figure*}[tb]
\includegraphics[width=\textwidth]{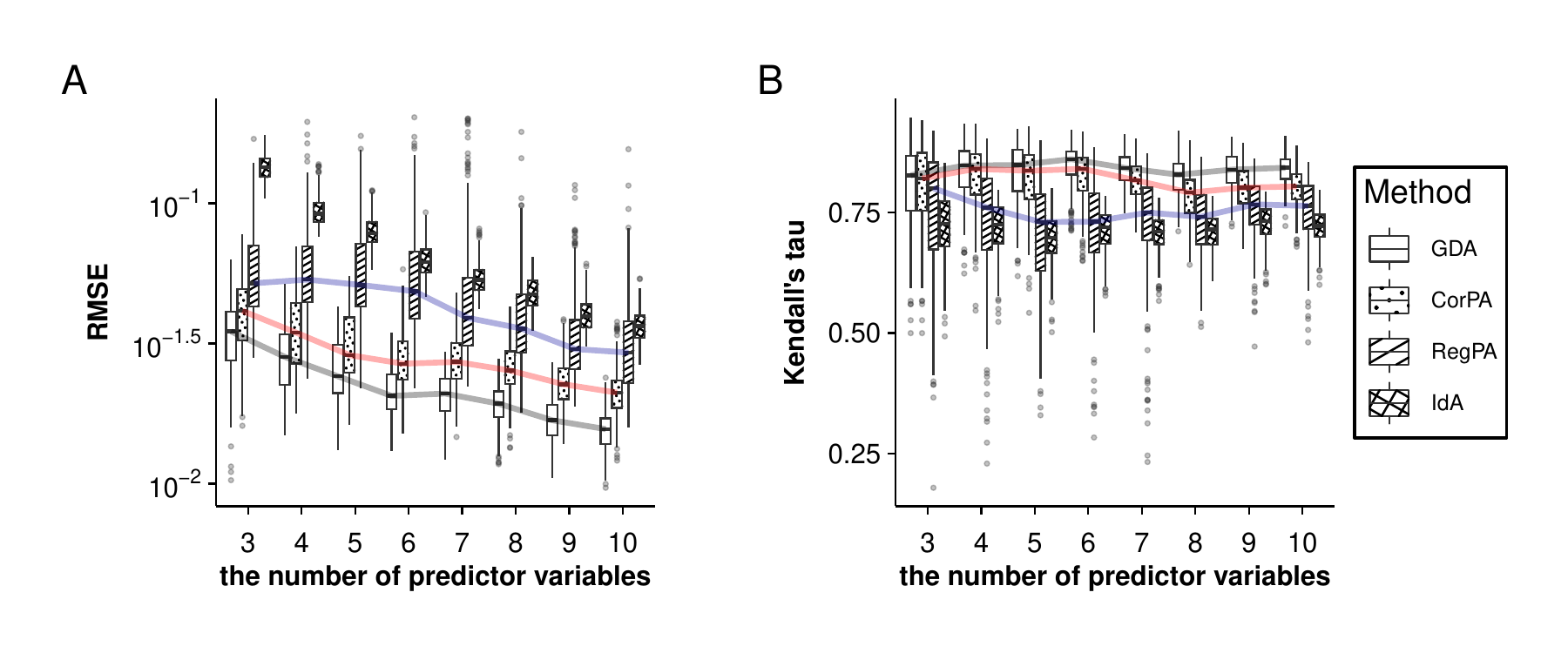}
\caption{Scenario 2.2 performance comparison of four reallocation methods using Johnson's $Z$ in ORMs, assessed by RMSE (left) and Kendall's tau (right). Red line indicates CorPA (RW)'s median performance, while blue and gray lines show RegPA (GCD) and GDA results, respectively.}
\label{fig: s2.2}
\end{figure*}

In Scenario 2.2 (Figure~\ref{fig: s2.2}), the performance of both $\mathrm{CorPA}^Z$ and $\mathrm{RegPA}^Z$ declines noticeably, with larger RMSE (and lower Kendall’s tau) values than $\mathrm{GDA}^Z$ across all $p$ though RW is still considered a better method than GCD.

\begin{figure*}[tb]
\includegraphics[width=\textwidth]{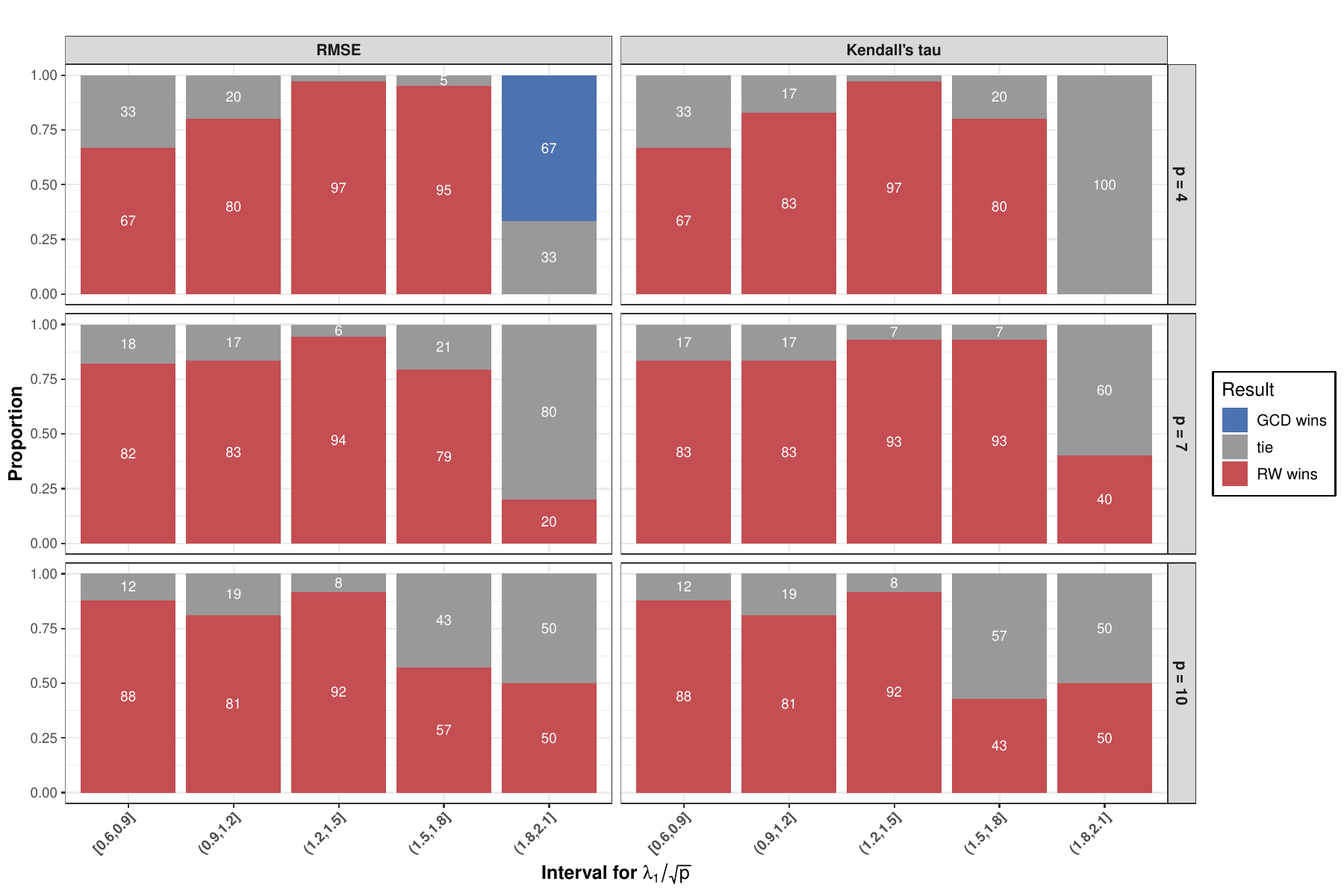}
\caption{Win-Loss Analysis between RW and GCD with severe multicollinearity (Scenario 2.1 and 2.2) in RMSE and Kendall’s tau.}
\label{fig: s2.1 and s2.2 win-loss}
\end{figure*}

The deterioration of $\mathrm{CorPA}^Z$ is attributed to its leveling problem, resulting from a strong first principal component. On the other hand, $\mathrm{RegPA}^Z$'s performance is still limited by a priori bias due to severe multicollinearity, i.e. a large $\mathrm{VIF}_{\max}/p$. Despite these limitations, the win-loss analysis in Figure~\ref{fig: s2.1 and s2.2 win-loss} confirms that $\mathrm{CorPA}^Z$ performs better than $\mathrm{RegPA}^Z$ in this scenario. We caution against the use of $\mathrm{CorPA}^Z$ (RW) due to its bias towards underestimating the importance of stronger predictors and overestimating that of weaker ones.

\section{Real-World Dataset Example}
In this section, we present our findings from real-world datasets, which corroborate the patterns observed in our simulations. We focus specifically on the transition from scenario 1.1 to 1.2, as this reveals critical distinctions in the performance of ORMs under different correlation structures.
\begin{itemize}
    \item Table~\ref{tab: real s1.1} showcases the results for the satisfaction with life dataset from~\cite{suh1998shifting}, previously analyzed by~\cite{azen2003dominance} to demonstrate the relative importance of five domain-specific predictors in predicting satisfaction scores $(n=428,p=5)$. This dataset exemplifies scenario 1.1, characterized by a mild first principal component $(\lambda_1/\sqrt{p}=0.88)$ and mild multicollinearity $(\mathrm{VIF}_{\max}/p=0.27)$. As anticipated, both RW and GCD perform similarly, with RW exhibiting a slightly better RMSE (RW: Kendall's $\tau = 1.00$, $\mathrm{RMSE} = 0.0013$; GCD: Kendall's $\tau = 1.00$, $\mathrm{RMSE} = 0.0016$).

    \begin{table}[t]
    \centering
    \caption{The relative importance analysis for satisfaction score in satisfaction with life dataset from~\cite{suh1998shifting}.}
    \label{tab: real s1.1}
        \begin{tabular}{llllll}
        \hline
         & Self  & Family & Finance & Housing & Health \\ \hline
         $\mathrm{GD}^\dagger$ & 50.51 & 25.08  & 11.53   & 9.05    & 3.84   \\ 
         $\mathrm{GCD}^\dagger$ & 51.09 & 25.12  & 11.26   & 8.99    & 3.53   \\ 
         $\mathrm{RW}^\dagger$ & 50.01 & 25.19  & 11.67   & 9.27    & 3.87   \\ 
         $\sum_{j=1}^p\mathrm{RegPA}_{ij}^Z$ & 1.01  & 1.00   & 1.00    & 1.00    & 1.00   \\ 
         VIF & 1.35  & 1.24   & 1.16    & 1.12    & 1.15   \\ 
         $\mathrm{GCD}^\dagger-\mathrm{GD}^\dagger$ & 0.59  & 0.04   & -0.26   & -0.06   & -0.31  \\ 
         $\mathrm{RW}^\dagger-\mathrm{GD}^\dagger$ & -0.50 & 0.11   & 0.14    & 0.22    & 0.04   \\ \hline
        \multicolumn{6}{l}{$\dagger$Normalized to sum 100\% $(R^2:0.50)$}                           \\ 
        \end{tabular}
    \end{table}
    
    \item Table~\ref{tab: real s1.2} demonstrates the job performance dataset from~\cite{johnson2001relative}, which utilizes seven performance dimension scores to predict overall performance ratings $(n=324,p=7)$, to represent scenario 1.2. This dataset exhibits a strong first principal component $(\lambda_1/\sqrt{p}=1.66)$ and mild multicollinearity $(\mathrm{VIF}_{\max}/p=0.50)$. The importance of the strongest predictor, NJSTP (Non-Job-Specific Task Proficiency), is noticeably underestimated, while that of the weaker predictors, such as WOCTP (Written and Oral Communication Task Proficiency), HWS (Handle Work Stress), and OCP (Organizational Citizenship Performance), are overestimated by RW due to its leveling problem under the strong first principal component. RW is clearly inferior to GCD in this case, as evidenced by the Kendall's $\tau=0.81$ and $\mathrm{RMSE} = 0.0074$. GCD, in contrast, appears to be a better choice with Kendall's $\tau = 0.90$ and $\mathrm{RMSE} = 0.0048$, aligning with the findings of this research but unexpected compared to the established findings in existing literature.

    \begin{table}[t]
      \caption{The relative importance analysis for overall performance rating in job performance dataset from~\cite{johnson2001relative}.}
      \label{tab: real s1.2}
      \centering
      \small 
      \setlength{\tabcolsep}{4pt}%
        \begin{tabular}{llllllll}
          \hline
                 & NJSTP & JSTP  & ICP   & JTC   & WOCTP & HWS   & OCP   \\ \hline
          $\mathrm{GD}^\dagger$          & 21.29 & 16.05 & 14.03 & 13.59 & 13.51 & 11.60 & 9.93  \\
          $\mathrm{GCD}^\dagger$         & 22.50 & 16.72 & 13.61 & 13.00 & 13.49 & 10.97 & 9.71  \\
          $\mathrm{RW}^\dagger$          & 19.05 & 16.95 & 13.82 & 13.61 & 14.14 & 11.83 & 10.60 \\
          $\sum_{j=1}^p\mathrm{RegPA}_{ij}^Z$ & 1.08  & 0.97  & 1.00  & 0.99  & 0.98  & 0.99  & 0.99  \\
          VIF         & 3.52  & 2.15  & 2.38  & 2.19  & 1.95  & 2.00  & 1.75  \\
          $\mathrm{GCD}^\dagger-\mathrm{GD}^\dagger$ & 1.20  & 0.68  & -0.42 & -0.59 & -0.03 & -0.62 & -0.22 \\
          $\mathrm{RW}^\dagger-\mathrm{GD}^\dagger$  & -2.24 & 0.90  & -0.20 & 0.02  & 0.62  & 0.23  & 0.68  \\ \hline
          \multicolumn{8}{l}{$\dagger$Normalized to sum 100\% $(R^2:0.75)$}
        \end{tabular}
    \end{table}

\end{itemize}
\section{Concluding Remarks}
Our comprehensive Monte Carlo simulations demonstrate that, among the orthogonalization methods considered, Johnson's $Z$ remains the most effective. For reallocation, we find that RW suffers from a leveling problem when the first principal component of the predictors is strong. Conversely, GCD is prone to an a priori bias when predictors exhibit severe multicollinearity. Both of these issues result in a decline in performance. Contrary to the widely held belief that RW is universally superior, our findings reveal that its effectiveness depends heavily on the correlation structure of the predictors. This analysis is possible due to the uniform sampling of a broad range of correlation structures in our simulation design.

Based on these insights, we propose the following guidelines for using ORMs: When the first principal component is mild, RW remains a suitable option. However, when the first principal component is strong, RW's performance declines due to the leveling problem. In such scenarios, GCD may offer a viable alternative if the multicollinearity is not severe, minimizing the risk of a priori bias. To assess the strength of the first principal component and severity of multicollinearity, one can refer to the threshold values provided in Table~\ref{tab: four scenarios}. If both the leveling problem and the a priori bias are likely to be substantial, we suggest that RW should be used with caution.

While these guidelines offer practical recommendations, further improvements in ORM performance may be possible. Our findings suggest that there may be orthogonalization methods superior to Johnson's $Z$ that have yet to be discovered. Furthermore, the consistent effectiveness of GDA suggests the potential for developing reallocation methods that could enhance ORM performance. Moreover, understanding the mechanisms by which ORMs approximate GD remains a challenging and intriguing question for future research.
\section*{Acknowledgment}
This work was partially supported by the Grant NSTC 106-2221-E-002-153-MY3 from National Science and Technology Council of Taiwan.
\appendix
\renewcommand{\thesection}{\Alph{section}}
\renewcommand{\thesubsection}{\Alph{section}.\arabic{subsection}}
\section{Appendix: Generating Response Variables with Desired Predictability}\label{app: A}
To simplify the generation process, we can focus on generating vectors on the unit sphere $S^{p-1}=\{u\in\mathbb{R}^p:\sum_{i=1}^p u_i^2 =1\}$ and then rescale the resulting response variable. This simplification is valid because relative importance measures, such as those used in our study, depend only on the correlation between the response variable and the predictors. To illustrate, given two vectors $u_1\in S^{p-1}$ and $u_2=0.5u_1$ share the same direction but differ in length, the resulting GDs and the ORMs will differ only by a constant factor of 0.5. This flexibility also allows us to adjust the coefficients of the explainable and unexplainable parts to achieve the desired $R_{y\cdot X}^2$.

For example, if we uniformly sample a vector $u$ from the unit sphere $S^{p-1}$ and desire a response variable with $R_{y\cdot X}^2=0.8$, as we did in this study, we can construct a response variable, $y'=Zu+0.5\epsilon$. The variable can then be normalized to unit length, $y=(Zu+0.5\epsilon)/\sqrt{1^2+0.5^2}$. We then have the correlations: $\rho_{zy}=Z^\top y=u/\sqrt{1^2+0.5^2}$. Given $\Sigma_{xz}=X^\top Z=VSV^\top$, the correlation between $X$ and $y$ can be derived as: $\rho_{xy}=X^\top y=X^\top Zu/\sqrt{1^2+0.5^2}=\Sigma_{xz} \rho_{zy}$. 

These calculations can be generalized to other orthogonalization techniques beyond Johnson’s $Z$, because they all span the same column space as the original predictors. For example, consider principal component, $Z_{\mathrm{PC}}$. We can write $Z_{\mathrm{PC}}=ZQ_{\mathrm{PC}}$, where $Q_{\mathrm{PC}}=V$ is an orthogonal matrix. With some algebraic manipulation, we can derive $\rho_{z_{\mathrm{PC}}y}=Z_{\mathrm{PC}}^\top y=Q_{\mathrm{PC}}^\top \rho_{zy}$ and $\Sigma_{xz_{\mathrm{PC}}}=X^\top Z_{\mathrm{PC}}=\Sigma_{xz} Q_{\mathrm{PC}}$. Similarly, we can use orthogonalization methods, such as Gram-Schmidt, where $Q_{\mathrm{GS}}$ is obtained by Cholesky decomposition of $\Sigma_{xx}$, or varimax, where $Q_{\mathrm{VM}}$ is obtained via varimax rotation.
\section{Appendix: Proofs}\label{app: B}
\subsection{\texorpdfstring{Preliminary Proof That Varimax is equivalent to Johnson’s $Z$ when $p=2$}{Preliminary Proof That Varimax is equivalent to Johnson’s Z when p=2}}
We begin by representing the predictors $x_1,x_2$ using the arbitrary orthogonal predictors $\tilde{z}_1,\tilde{z}_2$, which lie in the same subspace, such that
\begin{equation}
    \begin{cases}
        x_1 &= a\tilde{z}_1+b\tilde{z}_2 \\
        x_2 &= c\tilde{z}_1+d\tilde{z}_2 
    \end{cases}
    \tag{B1}
    \label{eqB1}
\end{equation}
where the predictors have unit norms $a^2+b^2=c^2+d^2=1$ and a correlation $\rho=ac+bd$. When Johnson’s $Z$ is applied, it can be shown that $a=d=1/2(\sqrt{1+\rho}+\sqrt{1-\rho})$ and $b=c=1/2(\sqrt{1+\rho}-\sqrt{1-\rho})$. The objective function of Varimax, as given in Equation~\eqref{eq: varimax}, can be written as
\begin{equation}
    O_{\mathrm{VM}}(\tilde{Z})=\frac{1}{2}\left((a^2-c^2)^2+(b^2-d^2)^2\right).
    \tag{B2}
    \label{eqB2}
\end{equation}
To maximize $O_{\mathrm{VM}}(\tilde{Z})$, the parameters $a,b,c$, and $d$ must satisfy the unit norm and correlation conditions stated above. As a result, the Lagrangian function of this constrained optimization is
\begin{align}
    O_{\mathrm{VM}}^L(\tilde{Z}) &=\frac{1}{2}\left((a^2-c^2)^2+(b^2-d^2)^2\right)+\lambda(ac+bd-\rho)+ \mu(a^2+b^2-1)+\gamma(c^2+d^2-1).
    \tag{B3}
    \label{eqB3}
\end{align}
where $\lambda, \mu, \gamma$ are Lagrange multipliers. By applying the first order conditions and substituting the values of $a,b,c,d$ from Johnson’s $Z$, we find that Johnson’s $Z$ is a local optimum (and potentially a global optimum) of $O_{\mathrm{VM}}(\tilde{Z})$. Although we can only demonstrate the local optimality, the equivalence between $Z_{\mathrm{VM}}$ and Johnson’s $Z$ can be visually confirmed through graphical tools and is empirically supported by numerical experiments. 

\subsection{Proof That the Expected Value of the ORMs and GD}
The ORM with an arbitrary orthogonalization and reallocation method can be written as follows:
\begin{equation}
    D_{A,i}^{\tilde{Z}}=\sum_{j=1}^p a_{ij}^{\tilde{Z}}\rho_{z_jy}^2,\quad\forall i,\ldots,p.
    \tag{B4}
    \label{eqB4}
\end{equation}
Assuming that $y$ is uniformly distributed on a sphere of radius $\sqrt{R^2_{y\cdot X}}$, we take the expectation on both sides. Due to the symmetry of the uniform distribution on a sphere, we have
\begin{equation}
    \mathbb{E}_y\left[D_{A,i}^{Z}\right]=\frac{R^2_{y\cdot X}}{p}\sum_{j=1}^p a_{ij}^Z,\quad\forall i=1,\ldots,p.
    \tag{B5}
    \label{eqB5}
\end{equation}
On the other hand, by introducing $\tilde{z}_1,\ldots,\tilde{z}_p$, GD can be rewritten as $D_{GDA}^{\tilde{Z}}$ along with some remaining terms:
\begin{equation}
    \mathrm{GD}(x_i)=D_{GDA}^{\tilde{Z}}+\sum_{j=1}^p\sum_{k>j}^p \rho_{\tilde{z}_j y}\rho_{\tilde{z}_k y}f_i(L_{\tilde{Z},j}, L_{\tilde{Z}, k}),
    \tag{B6}
    \label{eqB6}
\end{equation}
$\forall i=1,\ldots,p$, where $L_{\tilde{Z},j}$ is the $j$-th column of the loading matrix $L_{\tilde{Z}}$ and $f_i(L_{\tilde{Z},j}, L_{\tilde{Z}, k})$ is a function of correlation between $X$ and $\tilde{Z}$. Using the same steps as above and considering the row-sum of the $\mathrm{GDA}^{\tilde{Z}}$, we have
\begin{equation}
    \mathbb{E}_y[\mathrm{GD}(x_i)]=\frac{R^2_{y\cdot X}}{p},\quad \forall i=1,\ldots,p.
    \tag{B7}
    \label{eqB7}
\end{equation}

\subsection{\texorpdfstring{Proof That Row-sum of $\mathrm{GDA}^{\tilde{Z}}$ Equals to One}{Proof That Row-sum of GDA^(Z~) Equals to One}}

We begin with squared multiple correlation $R_{\tilde{z}_j\cdot X_S}^2$ of subset of predictors $X_S,S\subseteq P$ with $\tilde{z}_j$, which can be written as
\begin{equation}
    R_{\tilde{z}_j\cdot X_S}^2=\tilde{z}_j^\top X_S(X_S^\top X_S)^{-1}X_S^\top\tilde{z}_j=\tilde{z}_j^\top H_S\tilde{z}_j
    \tag{B8}
    \label{eqB8}
\end{equation}
where $H_S$ denotes the projection matrix onto the subspace spanned by $X_S$. Therefore, by summing the above equation from $i=1,\ldots,p$, we have 
\begin{align}
    \sum_{j=1}^p R_{\tilde{z}_j\cdot X_S}^2 &= \sum_{j=1}^p \tilde{z}_j^\top H_S\tilde{z}_j\notag\\
    &=\mathrm{Tr}\left(\tilde{Z}^\top H_S\tilde{Z}\right)\notag\\
    &=\mathrm{Tr}\left(\tilde{Z}\tilde{Z}^\top H_S\right)\notag\\
    &=\mathrm{Tr}\left(H_S\right)\notag\\
    &=|S|.
    \tag{B9}
    \label{eqB9}
\end{align}
Therefore, the i-th row-sum of the $\mathrm{GDA}^{\tilde{Z}}$ can be written as
\begin{equation}
\begin{aligned}
    \sum_{j=1}^p \mathrm{GD}_{\tilde{z}_j}(x_i) &= \sum_{j=1}^p \left(\frac{1}{p}\sum_{S\subseteq P\setminus\{i\}}\binom{p-1}{|S|}^{-1}\left(R_{\tilde{z}_j\cdot X_{S\cup\{i\}}}^2-R_{\tilde{z}_j\cdot X_{S}}^2\right)\right)\\[1mm]
    &=\frac{1}{p}\sum_{S\subseteq P\setminus\{i\}}\binom{p-1}{|S|}^{-1}\left(\sum_{j=1}^pR_{\tilde{z}_j\cdot X_{S\cup\{i\}}}^2-\sum_{j=1}^pR_{\tilde{z}_j\cdot X_{S}}^2\right)\\[1mm]
    &=\frac{1}{p}\sum_{S\subseteq P\setminus\{i\}}\binom{p-1}{|S|}^{-1}\left(|S\cup\{i\}|-|S|\right)\\[1mm]
    &=\frac{1}{p}\sum_{S\subseteq P\setminus\{i\}}\binom{p-1}{|S|}^{-1}\binom{p-1}{|S|}\\[1mm]
    &=\frac{1}{p}\times p = 1.
\end{aligned}
\tag{B10}\label{eqB10}
\end{equation}


\bibliographystyle{unsrtnat}
\bibliography{references}  

\end{document}